\documentclass[a4paper,11pt]{article}
\pdfoutput=1 
\usepackage{jheppub} 
\usepackage[T1]{fontenc} 
\usepackage{natbib}

\newtheorem{theorem}{Theorem}

\DeclareMathOperator{\Tr}{\mathrm{Tr}}
\newcommand{\AdS}{\textrm{AdS}}

\title{\boldmath Exact Stringy Microstates from Gauge Theories}

\author{Ji Hoon Lee}

\affiliation{Perimeter Institute for Theoretical Physics, Waterloo, Ontario, Canada N2L 2Y5}

\emailAdd{jihoon.lee@perimeterinstitute.ca}

\abstract{We study how the microstates of BPS sectors in string theory are organized in the dual $U(N)$ gauge theory. The microstates take the form of a coherent sum of stacks of branes and their open/closed string excitations. We propose a prescription to holographically construct the indices of string/brane configurations by analyzing the modifications of determinant operators in gauge theory. The string/brane configurations should be interpreted in the tensionless limit, but their indices are well-defined at finite $N$. In various examples, we provide evidence that a sum, of the giant graviton-type recently proposed in the literature, over all such configurations gives the finite $N$ gauge theory index. Finally, we discuss how these microstates assemble in the BPS Hilbert space and in what circumstances the branes can form bound states to produce black hole degeneracies.}

\begin{document} 
\maketitle
\newpage

\section{Introduction} \label{sec: introduction}

What are the microstates in string theory that underlie phenomena such as supersymmetric black holes? From the work of Strominger and Vafa \cite{Strominger:1996sh} and developments \cite{Callan:1996dv,Breckenridge:1996is,Sen:1995in,Ooguri:2004zv}, we know that the microstates take the form of configurations of strings and branes whose asymptotic degeneracies reproduce the entropy of BPS black holes. Still, there are many open questions that require a more precise understanding of string theory: How do the BPS microstates assemble to generate highly excited geometries? How do we know which string/brane configurations form a basis of states for the BPS Hilbert space? Can we enumerate such configurations?

Holography \cite{Maldacena:1997re,Witten:1998qj,Gubser:1998bc,Witten:1998zw}, in principle, can solve these problems by turning the counting of string and brane configurations into the counting of protected operators in a gauge theory \cite{Aharony:2003sx,Kinney:2005ej,Romelsberger:2005eg,Bhattacharya:2008zy}. In recent years, there have been many fruitful developments \cite{Benini:2015eyy,Cabo-Bizet:2018ehj,Choi:2018hmj,Benini:2018ywd,Honda:2019cio,ArabiArdehali:2019orz,Cabo-Bizet:2019eaf,ArabiArdehali:2019tdm,Kim:2019yrz,Cabo-Bizet:2019osg,Amariti:2019mgp,Lanir:2019abx,GonzalezLezcano:2019nca,Murthy:2020rbd,Agarwal:2020zwm,Benini:2020gjh,GonzalezLezcano:2020yeb,Copetti:2020dil,Cabo-Bizet:2020ewf,Cabo-Bizet:2020nkr,Goldstein:2020yvj,Jejjala:2021hlt,Aharony:2021zkr,Cassani:2021fyv,Choi:2021rxi,Choi:2021lbk,Boruch:2022tno} exploring the connection between superconformal indices of gauge theories and supersymmetric black holes in gravity duals. These works demonstrate that holographic field theories indeed contain the precise degrees of freedom to account for the degeneracies of higher-dimensional black holes.\footnote{See \cite{Gadde:2020yah} for an introduction to superconformal indices.}

Though the problem of counting degeneracies can be studied in gauge theory, a bulk understanding of string theory microstates has been absent. Clues regarding the structure of these bulk microstates have come from recent works \cite{Gaiotto:2021xce,Imamura:2021ytr} (see also \cite{Bourdier:2015wda,Arai:2019xmp,Arai:2019wgv,Arai:2019aou,Arai:2020qaj,Arai:2020uwd,Imamura:2021dya,Murthy:2022ien,Honda:2022hvy,Okazaki:2022sxo,Moosavian:2021ibw}) that found a reorganization of finite $N$ superconformal indices in terms of branes and their excitations in the string dual. In the work \cite{Gaiotto:2021xce} with D. Gaiotto, we proposed a brane expansion of the superconformal index by counting branes and their excitations via determinant operators in gauge theory. We review these results in Section \ref{subsec: counting open strings}. In the current work, we establish these developments in the setting of a general $U(N)$ gauge theory and explain the context in which the string/brane configurations that appear in the brane expansion should be understood as bulk microstates.

The aim of this work is to study how the microstates of BPS sectors in string theory are organized in the dual gauge theory. We propose that the microstates are organized as a coherent sum of branes that are dual to a set of determinant operators in gauge theory:
\begin{equation} \label{eq: general brane expansion}
    Z_N = Z_\infty \sum_{k_1,k_2, \cdots, k_s = 0}^\infty \left( x_1^{k_1 N} x_2^{k_2 N} \cdots x_s^{k_s N} \right) \ \hat{Z}_{(k_1,k_2, \cdots, k_s)}.
\end{equation}
$Z_N$ is the index of a $U(N)$ gauge theory or, via holography, the index of its string theory dual on a background with $N$ units of RR flux. There are $s$ ``types'' of branes wrapping different supersymmetric cycles. Terms in the sum are stacks of $k_i$ numbers of branes of the $i$-th type, where each brane has dimension/charges of order $N$. The brane index
\begin{equation}
    \hat{Z}_{(k_1,k_2, \cdots, k_s)}
\end{equation}
is that of the worldvolume $\prod_i U(k_i)$ quiver gauge theory, which enumerates the open string excitations on the $(k_1,k_2, \cdots, k_s)$ brane stack. The closed string spectrum $Z_\infty$ curiously factors out and has an interpretation in the supergravity regime as the index of Kaluza-Klein modes. Branes and their interactions can be summarized in a quiver diagram describing the field content of the worldvolume theory. One way to understand \eqref{eq: general brane expansion} is as a bulk-bulk duality between (1) coherent states of branes and their excitations in the stringy regime and (2) saddle geometries such as BPS black holes or bubbling geometries in the supergravity regime.

We provide a prescription to construct the indices $\hat{Z}_{(k_1,k_2, \cdots, k_s)}$ of worldvolume theories on bulk branes that are dual to a set of determinant operators in the boundary gauge theory. The input for the prescription is minimal: it requires only (1) a choice of a basis of charges to define the superconformal index and (2) a table of various quantum numbers of BPS fields. From this data, we can determine the integrand and the integration cycle for brane indices $\hat{Z}_{(k_1,k_2, \cdots, k_s)}$. We assume that $N$ is large during our derivation, but the surprise is that the final statement \eqref{eq: general brane expansion} seems to be an exact (i.e. convergent) statement at finite $N$.

The implications of our proposal concern the bulk, but our approach is purely gauge theoretic. Namely, our proposal is based on the counting of allowed modifications of determinant operators in a $U(N)$ gauge theory. In the spirit of 't Hooft, we will assume that any such gauge theory has a string theory dual. The string dual may or may not have a weakly-curved gravity regime. We can be agnostic about the brane interpretation of gauge theory determinants until we obtain the indices counting its modifications. If the dual string theory is known, we can interpret $\hat{Z}_{(k_1,k_2, \cdots, k_s)}$ as those counting open string excitations of certain branes in the theory. Indeed, in situations where AdS/CFT correspondence is well-established, our ``brane'' indices agree with direct bulk computations that count modes on branes wrapped on supersymmetric cycles in the probe-brane approximation \cite{Imamura:2021ytr}. In situations where the string theory dual is less transparent, our results can be taken as novel predictions for the (indices of) string/brane configurations.

Let us comment on the parameter regime in which the microstates should be interpreted, starting at large $N$. At large $N$, the microstates should be thought as string/brane configurations in the tensionless limit of zero 't Hooft coupling $\lambda$, as opposed to the supergravity regime of large $\lambda$. The reason is that the branes indices are computed in a free $U(N)$ gauge theory, so the proposed branes are dual to determinants of free fields. In gauge theory, turning on $\lambda$ even slightly can introduce nontrivial $Q$-cohomology in the counting of BPS operators \cite{Aharony:2003sx,Kinney:2005ej,Chang:2013fba}. So a determinant operator may no longer sit inside the cohomology at $\lambda > 0$. Nonetheless, brane indices should be computable at any value of the 't Hooft coupling though their interpretation at finite $\lambda$ will be less accurate.

The tensionless regime provides the correct context for interpretation also because we are proposing that the string/brane configurations, whose indices we construct, are indeed microstates. In the limit of large $\lambda$ and charges, BPS configurations of strings and branes should \textit{become} supergravity solutions such as BPS black holes \cite{Gutowski:2004ez,Gutowski:2004yv,Chong:2005hr,Kunduri:2006ek} or bubbling geometries \cite{Lin:2004nb,Gaiotto:2009gz,Bena:2007kg,Lunin:2002iz}. Therefore, the microstates should not be thought as sitting inside a supergravity solution. The tensionless point is an optimal place to present the string theory microstates because we can be certain that there are no additional gravitational contributions such as those from saddle geometries.

At finite $N$, the interpretation of $\hat{Z}_{(k_1,k_2, \cdots, k_s)}$ as open string excitations on a stack of branes breaks down. Despite the lack of available interpretation, brane indices are well-defined for any value of $N$. In fact, they are independent of $N$ apart from bare fugacity prefactors $x_i^{k_i N}$ with trivial $N$-dependence. At small $N$, the only tractable description is as determinant operators in the dual $U(N)$ gauge theory, but even here the exact operator corresponding to a ``modified determinant'' in gauge theory is far from clear. This is because there is a nontrivial analytic continuation in the counting of determinant modifications which we explain in detail later. Determinant operators seem to be a good starting point for constructing microstates in the string dual, but they do not appear to play any additional role in the finite $N$ $Q$-cohomology of gauge theory due to the analytic continuation.

In Section \ref{sec: organizing principles}, we begin with a review of the counting of open string excitations. The spectrum of these excitations requires an analytic continuation. We present a precription to implement the analytic continuation based on physical modifications of determinant operators in gauge theory, thereby constructing the index of string/brane configurations. Then we propose how these microstates assemble in string duals of $U(N)$ gauge theories.

In Section \ref{sec: examples}, we demonstrate our prescription in physical examples. We find the appearance of a wide class of gauge theory determinants, such as those consisting of fermion bilinears or of field strength components. This finding enables us to present the microstate configurations in string duals of $\mathcal{N}=1$ and $\mathcal{N}=2$ vectormultiplets. We discuss a puzzle in the $1/16$-BPS sector of $\mathcal{N}=4$ SYM.

In Section \ref{sec: structure of bulk microstates}, we study how the microstates assemble in the BPS Hilbert space in a manner consistent with gauge theory and supergravity. We argue that the bulk Hilbert space acquires an effective grading corresponding to the RR charges of bulk branes. We explain when the branes can form bound states to produce BPS black holes or bubbling geometries. We conclude with a list of open questions in Section \ref{sec: discussion}.

Appendix \ref{app: derivation half-bps} contains an explicit derivation of brane indices for the half-BPS sector. Appendix \ref{app: multivariate residues} is a brief review of the theory and computation of multivariate residues. In Appendix \ref{app: m2 higgs branch}, we apply our prescription to the Higgs sector of the M2 worldvolume theory, an example with (anti)fundamental fields. In Appendix \ref{app: n1 yang mills checks} and \ref{app: n2 yang mills checks}, we provide checks of our proposal for theories consisting of $\mathcal{N}=1$ and $\mathcal{N}=2$ vectormultiplets.

\section{BPS microstates in string theory} \label{sec: organizing principles}

\subsection{Counting open string excitations}  \label{subsec: counting open strings}

We begin by reviewing the prescription of \cite{Gaiotto:2021xce} to ``count'' at large $N$ the operators in a $U(N)$ gauge theory which could be obtained as modifications of products of determinants
\begin{equation} \label{eq: determinant products}
    (\det X_1)^{k_1} (\det X_2)^{k_2} \cdots (\det X_s)^{k_s}
\end{equation}
for a set of adjoint fields or letters $X_i$. The basic strategy was to fermionize the determinants by adding some auxiliary zero-dimensional (anti)fundamental fermionic degrees of freedom, together with some bosonic antifields to account for the Ward identities of the auxiliary fermion integrals. Some fermionic zeromodes had to be removed by hand at the end of the calculation.

The effective theory of excitations of determinants took the form of a $\prod_i U(k_i)$ quiver gauge theory with fields counted by explicit single-letter indices derived from those of the $U(N)$ gauge theory. In all holographic examples we checked, the single-letter indices of the quiver gauge theory coincide with those of the worldvolume theories of certain giant graviton branes which are expected to be dual to the determinant operators. 

Consider a $U(N)$ gauge theory with adjoint and (anti)fundamental fields. Adjoint fields are counted by the single-letter index $f$ and (anti)fundamental fields are counted by $v,\bar{v}$. Concretely, this means that the index for the gauge theory is written as an integral over $U(N)$ fugacities $\mu_a$:
\begin{equation} \label{eq: gauge theory index with fund}
    Z_N = \frac{1}{N!} \oint_{|\mu_a|=1} \prod_{a=1}^{N} \frac{d\mu_a}{2\pi i \mu_a} \prod_{a \neq b} (1 - \mu_a \mu_b^{-1} ) \ \mathrm{PE} \left[ f \sum_{a,b} \mu_a \mu_b^{-1} + v \sum_{a} \mu_a + \bar{v} \sum_b \mu_b^{-1} \right],
\end{equation}
where $\mathrm{PE}$ denotes the plethystic exponential.\footnote{The plethystic exponential of a function $h$ is defined as \[ \mathrm{PE}[h(y_j) ] = \exp \left( \sum_{n=1}^\infty \frac{1}{n} h(y_j^n) \right). \] By a formal exchange of sums, it can be expressed as a (infinite) product in the $U(N)$ integrand of a superconformal index.} At large $N$, the index becomes
\begin{equation} \label{eq: Zinf with fund}
    Z_\infty = \mathrm{PE}\left[ \frac{v \bar{v}}{1-f} \right] \prod_{n=1}^\infty \frac{1}{1 - f(y_A^n)},
\end{equation}
where $y_A$ are fugacities associated with global symmetry charges of the $U(N)$ gauge theory. This expression has a transparent interpretation. The denominator counts polynomials in traces built from adjoint fields. The numerator counts polynomials in mesonic operators, built from an arbitrary number of adjoint fields sandwiched between an anti-fundamental field and a fundamental field.

For simplicity, we limit our examples in the main text to the case of $U(N)$ gauge theories with only adjoint fields. Still, the expression \eqref{eq: Zinf with fund} will be useful shortly. We treat in detail an example containing (anti)fundamental fields in Appendix \ref{app: m2 higgs branch}. With only adjoint fields, i.e. $v=\bar{v}=0$, the $U(N)$ gauge theory index becomes
\begin{equation} \label{eq: gauge theory index}
    Z_N = \frac{1}{N!} \oint_{|\mu_a|=1} \prod_{a=1}^{N} \frac{d\mu_a}{2\pi i \mu_a} \prod_{a \neq b} (1 - \mu_a \mu_b^{-1} ) \ \mathrm{PE} \left[ f \sum_{a,b} \mu_a \mu_b^{-1} \right].
\end{equation}
At large $N$, we have
\begin{equation} \label{eq: Zinf}
    Z_\infty = \prod_{n=1}^\infty \frac{1}{1 - f(y_A^n)}.
\end{equation}

The prescription developed in \cite{Gaiotto:2021xce} to count modifications of products of determinants \eqref{eq: determinant products} involves three steps: (1) we add $k_i$ copies of a system of auxiliary (anti)fundamental fermionic and bosonic fields for each type of determinant, (2) we take the large $N$ limit, and then (3) we project onto $\prod_i U(k_i)$ invariants. The original $U(N)$ gauge theory has vanishing $v,\bar{v}$, but the auxiliary (anti)fundamental fields in the theory with determinant insertions now has nonzero (anti)fundamental single-letter indices $v, \bar{v}$. The auxiliary degrees of freedom have single-letter indices
\begin{align}
    v^i &= (x_i - 1) \sigma^{X_i} \\
    \bar{v}_j &= (1 - x_j^{-1})/ \sigma^{X_j}.
\end{align}
where $x_i$ is the fugacity for $X_i$. There is freedom to introduce additional variables $\sigma^{X_i}$ which become gauge fugacities for the effective quiver gauge theory.

At large $N$, the index of the $U(N)$ gauge theory in the presence of determinants will look like \eqref{eq: Zinf with fund}. There is now an ``effective'' single-letter index
\begin{equation}
    \frac{v^i \bar{v}_j}{1 - f},
\end{equation}
on top of the polynomials of traces in the original gauge theory. The single-letter index describing the interactions between determinants of types $i$ and $j$ is
\begin{equation}
    \hat{f}^i_j = \delta^i_j + \frac{(x_i - 1)(1 - x_j^{-1})}{1 - f}
\end{equation}
Adjoint excitations $i = j$ correspond to open strings that end on the same type of brane. Bifundamental excitations $i \neq j$ are open strings that start at a brane of type $i$ and end on a brane of type $j$. The $\delta^i_j$ factor cancels fermion zeromodes that occur for adjoint interactions.\footnote{If (anti)fundamental indices $v, \bar{v}$ were nonzero in the original theory, we would also have
\begin{equation*}
    \hat{v}^i = \frac{(x_i - 1)}{1-f} v , \qquad \hat{\bar{v}}_j = \frac{(1 - x_j^{-1})}{1-f} \bar{v}.
\end{equation*}}

Let us treat the extra variables $\sigma^{X_i}$ as $U(k_i)$ gauge fugacities and impose gauge invariance for all brane types $i$. The result is the index of an effective $\prod_i U(k_i)$ quiver gauge theory
\begin{align} \label{eq: general brane index}
    \hat{Z}_{(k_1, \cdots, k_s)} = &\frac{1}{k_1! \cdots k_s!} \oint \prod_{{a_1}=1}^{k_1} \frac{d\sigma^{X_1}_{a_1}}{2\pi i \sigma^{X_1}_{a_1}} \cdots \prod_{{a_s}=1}^{k_s} \frac{d\sigma^{X_s}_{a_s}}{2\pi i \sigma^{X_s}_{a_s}} \nonumber \\
    &\prod_{a_1 \neq b_1} \left( 1 - \frac{\sigma^{X_1}_{a_1}}{\sigma^{X_1}_{b_1}} \right) \cdots \prod_{a_s \neq b_s} \left( 1 - \frac{\sigma^{X_s}_{a_s}}{\sigma^{X_s}_{b_s}} \right) \mathrm{PE} \left[ \sum_{i,j=1}^s \left( \hat{f}^i_j \sum_{a_i,b_j} \frac{ \sigma^{X_i}_{a_i} }{ \sigma^{X_j}_{b_j} } \right) \right],
\end{align}
whose holographic interpretation is the worldvolume gauge theory on stacks of branes in the string dual of the $U(N)$ gauge theory. The brane index $\hat{Z}_{(k_1, \cdots, k_s)}$ is accompanied by bare fugacities
\begin{equation}
    x_1^{k_1 N} \cdots x_s^{k_s N}
\end{equation}
held by a collection of determinant operators \eqref{eq: determinant products} and by the spectrum $Z_\infty$ of background closed strings \eqref{eq: Zinf}. We have not yet specified the integration cycle. The correct cycle is nontrivial and will be the subject of the next subsection.

Our review of previous results was rather schematic. For completeness, we include an explicit derivation of brane indices for the half-BPS sector in Appendix \ref{app: derivation half-bps}.

\subsection{Analytic continuation of the counting} \label{subsec: analytic continuation}

An important feature of the worldvolume theory is that it contains ``fields'' which have charges opposite to the ones of the original theory. For example, when counting modifications of $\det X$ we have to include modifications where some $X$'s in the determinant are replaced by the identity matrix $I$. This modification has an effective charge which is opposite to the one of $X$. This can create a problem: if the charges of the effective fields do not lie in a convex cone, infinitely many operators can contribute in any given charge sector and the operator counting problem is ill-defined. In all situations we work in, various cancellations ensure that the modifications of powers of a single determinant give rise to a well-defined counting problem, giving the index of the worldvolume theory on stacks of various branes.

When considering modifications of multiple determinants, the problem becomes more severe. Consider, for example, modifications of $\det X \det Y \propto$
\begin{equation}
    \left( \epsilon_{i_1 i_2 \cdots i_N} \epsilon^{j_1 j_2 \cdots j_N} \ X^{i_1}_{j_1} X^{i_2}_{j_2} \cdots X^{i_N}_{j_N} \right) \left( \epsilon_{k_1 k_2 \cdots k_N} \epsilon^{l_1 l_2 \cdots l_N} \ Y^{k_1}_{l_1} Y^{k_2}_{l_2} \cdots Y^{k_N}_{l_N} \right).
\end{equation}
A replacement of a single $X \to Y$ has precisely opposite charge to $Y \to X$. Mixed modifications will compound the problem. From the point of view of branes, the problem is that we now have configurations of branes wrapping transverse directions: fields which describe movement of a brane in a direction parallel to a second brane have opposite charge to the derivatives of fields on the worldvolume of the latter. This obstructs a standard counting of operators or excitations. 

Ultimately, we are interested in what could be called an analytic continuation of a counting problem. Consider again the case discussed in the previous work \cite{Gaiotto:2021xce}. The objective was to compare the counting of determinant modifications with the finite $N$ corrections to the index of the $U(N)$ gauge theory. Denote with $x$ the fugacity of field $X$ in the original theory. The index of the theory can be interpreted as a formal power series in $x$ or as an actual function of $x$, defined initially for sufficiently small $x$. The proposal there was to compute the index $\tilde{Z}_k$ for modifications of $(\det X)^k$ as a power series in $x^{-1}$ and then analytically continue its resummation to a power series $\hat{Z}_k$ in $x$. This analytically-continued brane index could then be identified as finite $N$ corrections to the index of the original theory.

In the case of powers $\prod_i (\det X_i)^{k_i}$ of multiple determinants, we cannot directly define the analogue of $\tilde{Z}_k$. Instead, we need to find a direct definition of the analytically-continued brane index $\hat{Z}_{(k_1, k_2, \cdots)}$ that can be identified with finite $N$ corrections to the gauge theory index. Our strategy is to define $\hat Z_{(k_1, k_2, \cdots)}$ by evaluating the gauge fugacity integral along an integration cycle, defined implicitly through the theory of multivariate residues \cite{griffiths2014principles,Larsen:2017aqb}. The integration cycle will directly reflect the set of allowed open string excitations of the brane stack. Another integration cycle was discussed in \cite{Imamura:2021ytr}. Our prescription gives a physical explanation for the integration cycle and generalizes the prescription to cycles for determinants in general $U(N)$ gauge theories.

Concisely, a multivariate residue is defined by factoring the denominator of an integrand
\begin{equation}
    \frac{h(\sigma) d\sigma_1 \wedge \cdots \wedge d\sigma_K}{g_1(\sigma) \dots g_K(\sigma)},
\end{equation}
into $K = \sum_i k_i$ parts. The residue is evaluated at an isolated common zero of the $g_a$, which for us will be the point $\sigma=0$ where all $\prod_i U(k_i)$ fugacities vanish. The multivariate residue selects a specific integration cycle in the neighborhood of the zero: the torus $|g_a(\sigma)| =\epsilon$ for sufficiently small $\epsilon$, oriented according to the condition $d(\arg g_1) \wedge \cdots \wedge d(\arg g_K) \geq 0$. The integration cycle and the final answer depend on the choice of the denominator partition\footnote{We inserted parentheses in \eqref{eq: denominator partition} to guide the viewing of the expression. The set $g$ has $K = \sum_i k_i$ elements and should be understood without the parentheses.}
\begin{equation} \label{eq: denominator partition}
    g = \Big\{ \big( g_{1}^{X_1}, \cdots, g_{k_1}^{X_1} \big), \big( g_{1}^{X_2}, \cdots, g_{k_2}^{X_2} \big), \cdots, \big( g_{1}^{X_s}, \cdots, g_{k_s}^{X_s} \big) \Big\},
\end{equation}
so the task is to find $g$ that will implement the correct analytic continuation of the brane index. 

Roughly speaking, our prescription for $g$ is that $g_{a_i}^{X_i}$ contains denominators that each represent single open strings ending on the brane $(\det X_i)_{a_i}$, where we now ``number'' the determinants of \eqref{eq: determinant products} as
\begin{equation} \label{eq: determinant products partitioned}
    \Big( (\det X_1)_1 \cdots (\det X_1)_{k_1} \Big) \Big( (\det X_2)_1 \cdots (\det X_2)_{k_2} \Big) \cdots \Big( (\det X_s)_1 \cdots (\det X_s)_{k_s} \Big).
\end{equation}
The analytically-continued spectrum of multiple open strings can be read off from the final brane index defined in terms of $g$. We will make precise the above statements shortly. We review the theory and computation of multivariate residues in Appendix \ref{app: multivariate residues}.

The integrand of the brane index \eqref{eq: general brane index} generally involves ratios of infinite products. Numerators and denominators come respectively from negative and positive terms that appear in an expansion of the brane single-letter index $\hat{f}^i_j$. If we have
\begin{equation}
    \hat{f}^i_j (y_A) = +\sum_\alpha p_\alpha(y_A) - \sum_\beta n_\beta(y_A)
\end{equation}
for some $i$ and $j$, the denominator involves products of
\begin{equation} \label{eq: product element example}
    \left( \sigma_{a_i}^{X_i} - p_\alpha (y_A) \sigma_{b_j}^{X_j} \right),
\end{equation}
where $p_\alpha(y_A)$ is some product of global symmetry fugacities $y_A$ and their inverses. If $p_\alpha(y_A)$ is the fugacity associated with a modification of $(\det X_i)_{a_i}$ by explicit letters in the original gauge theory, then we can interpret the denominator \eqref{eq: product element example} as a single open string that starts on brane $(\det X_j)_{b_j}$ and ends on brane $(\det X_i)_{a_i}$. We will take this interpretation in a literal manner in the context of gauge/string duality. The labels on gauge fugacities $\sigma$ are the Chan-Paton labels of the open string endpoints. On the brane $(\det X_i)_{a_i}$ where it ends, the string sources charges that can be read off from the fugacity $p_\alpha(y_A)$.

We will say that a fugacity corresponds to an \textit{allowed} modification if there exists a replacement, carrying the same fugacity, of a single letter $X_i$ in $\det X_i$ by an explicit collection of BPS fields in the $U(N)$ gauge theory. Consider the denominator above, which we can also write as
\begin{equation} \label{eq: denominator ambiguity}
    \left( \sigma_{a_i}^{X_i} - p_\alpha (y_A) \sigma_{b_j}^{X_j} \right) \ \propto \ \left( \sigma_{b_j}^{X_j} - p^{-1}_\alpha (y_A) \sigma_{a_i}^{X_i} \right).
\end{equation}
In many situations, fugacity $p^{-1}_\alpha(y_A)$ will not correspond to any allowed modification of $(\det X_j)_{b_j}$. On the other hand, fugacity $p_\alpha(y_A)$ always corresponds to an allowed modification of $(\det X_i)_{a_i}$. So in these situations, the denominator induces a preferred orientation on the open string.

However, there are circumstances where $p_\alpha^{-1}(y_A)$ corresponds to an allowed modification of $(\det X_j)_{b_j}$ too. In other words, it can happen that both the left and right hand sides of \eqref{eq: denominator ambiguity} admit interpretations as open strings related only by orientation reversal and charge conjugation. It turns out that the value of the residue can differ based on whether we classify the denominator into $g_{a_i}^{X_i}$ or $g_{b_j}^{X_j}$. This is because we are making a choice on the orientation of this string while keeping the orientations of all other open strings fixed. To get an invariant answer, we introduce an ordering for the branes. A convenient ordering is written in \eqref{eq: denominator partition} and \eqref{eq: determinant products partitioned}. It suggests that we first partition into $g_1^{X_1}$ all denominators that can represent open strings ending on $(\det X_1)_1$. We then partition into $g_2^{X_1}$ all leftover denominators that can represent open strings ending on $(\det X_1)_2$, and repeat the process until $(\det X_s)_{k_s}$. This means that if $g_{a_i}^{X_i}$ comes before $g_{b_j}^{X_j}$ in the ordering, then we place \eqref{eq: denominator ambiguity} into $g_{a_i}^{X_i}$. Vice versa holds if $g_{b_j}^{X_j}$ comes before $g_{a_i}^{X_i}$. The final result does not depend on the brane ordering.

Our analysis of the allowed modifications points to a natural prescription for the denominator partition $g$: Let $g$ be ordered as
\begin{equation} 
    g = \Big\{ \big( g_{1}^{X_1}, \cdots, g_{k_1}^{X_1} \big), \big( g_{1}^{X_2}, \cdots, g_{k_2}^{X_2} \big), \cdots, \big( g_{1}^{X_s}, \cdots, g_{k_s}^{X_s} \big) \Big\}
\end{equation}
and gauge fugacities $\sigma$ also ordered accordingly. 
The prescription for $g$ is that $g_{a_i}^{X_i}$ contains denominators that represent allowed modifications ending on the brane $(\det X_i)_{a_i}$, where denominators are partitioned into elements of $g$ in the above order. Elements $g_{a_i}^{X_i}$ should also include powers of the corresponding zeromode $\sigma_{a_i}^{X_i}$ that appear in the denominator of the integrand. Elements of $g$ have a common degenerate zero $g=0$ at the point $\sigma=0$ where all gauge fugacities vanish. We claim that the residue of the $\prod_i U(k_i)$ gauge integral at $\sigma=0$, with the integration cycle given by $g$, yields the brane index $\hat{Z}_{(k_1,k_2, \cdots, k_s)}$ that can be compared to the gauge theory spectrum.

We demonstrate our prescription in a series of examples in Section \ref{sec: examples}.

\subsection{Structure of microstates} \label{subsec: structure of microstates}

We are ready to address how BPS microstates in string theory are organized in the dual gauge theory. We propose that string theory microstates are organized as a coherent sum of branes that are dual to a set of determinant operators in gauge theory:
\begin{equation} \label{eq: general brane expansion 2}
    Z_N = Z_\infty \sum_{k_1,k_2, \cdots, k_s = 0}^\infty \left( x_1^{k_1 N} x_2^{k_2 N} \cdots x_s^{k_s N} \right) \ \hat{Z}_{(k_1,k_2, \cdots, k_s)}.
\end{equation}
The ingredients in \eqref{eq: general brane expansion 2} can be derived in a free $U(N)$ gauge theory: $Z_N$ is the finite $N$ gauge theory index, $x_i^{k_i N}$ are bare fugacities of determinants $(\det X_i)^{k_i}$, and $\hat{Z}_{(k_1,k_2, \cdots, k_s)}$ is an analytic continuation of the counting of determinant modifications. However, the most natural setting in which to interpret \eqref{eq: general brane expansion 2} is the string theory dual. By gauge/string duality, $Z_N$ is the finite $N$ string theory index, $Z_\infty$ is the spectrum of closed strings, $x_i^{k_i N}$ are fugacities of branes dual to determinants, and $\hat{Z}_{(k_1,k_2, \cdots, k_s)}$ is an analytic continuation of the spectrum of open string excitations on stacks of branes. The equation says that the BPS index of finite $N$ string theory is exactly (i.e. convergently) the sum of all brane/string configurations, where the sum is over numbers of branes. The proposal for the microstates in various examples will be based on the set of gauge theory determinant operators that contribute to \eqref{eq: general brane expansion 2}, computed via the prescription provided in the current section.

\begin{figure}[t]
\centering
\includegraphics[scale=0.35]{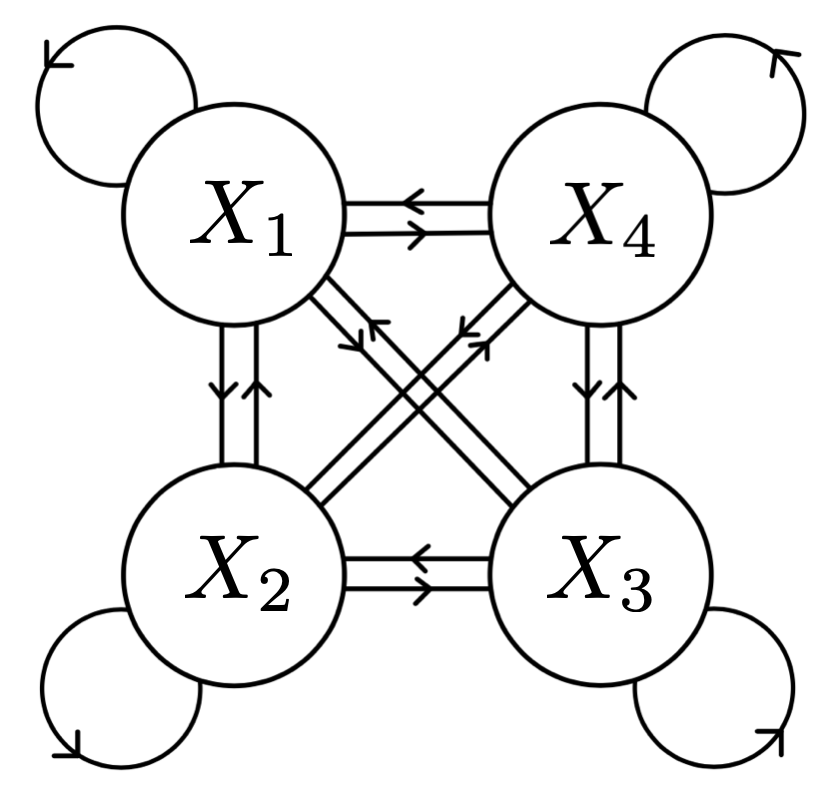}
\caption{General quiver diagram for $s=4$.}
\label{fig: general quiver}
\end{figure}

Branes and their interactions can be captured in a quiver diagram as in Figure \ref{fig: general quiver}. The quiver diagram is that of the $\prod_i U(k_i)$ quiver gauge theory that lives on the worldvolume of intersecting branes. Each node is associated with a brane dual of $(\det X_i)^{k_i}$ where there are $k_i$ coincident branes of the $i$-th type. In the main text, we consider quivers with adjoint and bifundamental arrows. Adjoint arrows start and end on the same node, and they represent open strings that start and end on the same type of brane. Bifundamental arrows start from one node and end on a different node, and they represent open strings that start on one type of brane and end on a different type of brane. In general, nodes are maximally connected by all-to-all arrows. The brane single-letter index $f^i_j$ is a matrix that encodes single-letter excitations associated with each arrow. If the original gauge theory has (anti)fundamental fields, the brane quiver can also have arrows that start (end) on a node but do not end (start) anywhere. We work through an example with (anti)fundamentals in Appendix \ref{app: m2 higgs branch}.

BPS free fields in the $U(N)$ gauge theory rearrange themselves and play different roles in the effective theory of determinant modifications. For example, a bosonic scalar or an adjoint fermion pair may assume the role of a derivative on the worldvolume theory interpreted at large $N$. This is a general phenomenon because the arrangement of branes in relation to one another affects charges that can be sourced on the worldvolume by open strings.

Determinant operators whose modifications make up the finite $N$ spectrum are intriguing in their own right. In explicit examples, we find the appearance of not only determinants of scalar fields (e.g. $\det X$), which were studied in the context of giant graviton branes \cite{McGreevy:2000cw,Grisaru:2000zn,Hashimoto:2000zp,Balasubramanian:2001nh,Mikhailov:2000ya,Biswas:2006tj,Mandal:2006tk,Kim:2006he,Corley:2001zk,Berenstein:2005aa,Balasubramanian:2004nb,deMelloKoch:2007rqf,Yang:2021kot,Jiang:2019xdz,Budzik:2021fyh}, but also the determinant $\det F$ of a field strength component and heavy fermion condensates such as a gaugino-gaugino pair $\det \lambda_+ \lambda_-$ and a gaugino-chiralino pair $\det \lambda_\pm \psi$.

Our proposal for the microstates is based on gauge theory determinant operators that are necessary in the brane expansion \eqref{eq: general brane expansion 2} of the superconformal index. It is unclear whether the proposed microstates (whose indices we compute) actually form a basis of the free BPS Hilbert space of finite $N$ gauge/string theory. It is possible that some branes are cancelled in the index. It would be important to investigate whether the microstates form a basis by formulating a brane expansion for the free BPS partition function. We nonetheless expect that the proposed microstates of Section \ref{sec: examples} are sufficient to capture the degeneracies of BPS black holes or bubbling geometries at large $N$.

It is natural to wonder whether the ``basis'' of microstates suggested by our prescription is unique. More precisely, we should ask whether another set of string/brane indices can assemble as in \eqref{eq: general brane expansion 2} to produce the same finite $N$ index spectrum. We believe there are other equally-valid sets of string/brane indices. Indeed, if we change the basis of charges in the definition of the gauge theory index, the set of contributing determinants also seems to change. From the perspective of string theory, it is not surprising that seemingly different branes correspond to the same set of states \cite{Myers:1999ps}. The microstates proposed in Section \ref{sec: examples} are those that are given by our prescription for a convenient basis of global symmetry charges $\mathcal{C}_A$ in gauge theory. An interesting direction would be to study how the set of contributing branes vary with different sets of charges $\mathcal{C}_A$ that commute with the supercharge $Q$.

\section{Examples} \label{sec: examples}

In this section, we demonstrate our prescription to construct the indices of string theory microstates. The set of branes that appear in $1/2$- and $1/4$-BPS sectors of IIB string theory on $\AdS_5 \times S^5$ are giant gravitons dual to determinants of scalar fields. The example of a $U(N)$-gauged fermion is somewhat surprising because it shows that there are branes dual to determinants of fermion bilinears. These branes play an essential role in the brane expansion, but they are not giant gravitons in the standard sense. Lessons from the fermion example enables us to propose brane configurations in string duals of holomorphically-twisted $\mathcal{N}=1$ and $\mathcal{N}=2$ gauge theories. We end with a puzzle in the $1/16$-BPS sector regarding other determinants that can show up in the brane expansion.

\subsection{Half-BPS sector} \label{subsec: half-bps}

We begin with the simplest example of the half-BPS index of $U(N)$ $\mathcal{N}=4$ super Yang-Mills. The half-BPS sector in gauge theory is spanned by multi-trace operators of the form $\prod_i \Tr(X^{m_i})$ modulo trace relations at finite $N$. The field $X$ is a bosonic adjoint scalar that has charge $1$ under R-symmetry $U(1)_X \subset SO(6)$. There are no fermions in this sector, so the index is in fact the half-BPS partition function.

The gauge theory single-letter index in the half-BPS sector is $f = x$. In this example, the finite $N$ index can be written exactly:
\begin{equation}
    Z_N = \prod_{n=1}^N \frac{1}{1 - x^n}.
\end{equation}
In \cite{Gaiotto:2021xce}, it was proposed that $Z_N$ can be reorganized as a sum over terms that are readily interpreted as bulk giant gravitons and their open string excitations:
\begin{equation} \label{eq: 1/2-BPS GGE}
    Z_N = Z_\infty \sum_{k=0}^\infty x^{k N} \hat{Z}_k
\end{equation}
where
\begin{equation} \label{eq: half-BPS hatZ}
    \hat{Z}_k = \prod_{m=1}^k \frac{1}{1 - x^{-m}}
\end{equation}
is the index of the worldvolume gauge theory on $k$ coincident giant gravitons and where
\begin{equation}
    Z_\infty = \prod_{n=1}^\infty \frac{1}{1 - x^{n}}
\end{equation}
represents background closed strings. The giant gravitons wrap a maximal $S^3 \subset S^5$ that is fixed under the $U(1)_X$ isometry.

This proposal was based on the prescription described in Section \ref{subsec: counting open strings}, which says that the adjoint single-letter index of the giant graviton worldvolume theory is $\hat{f}^{X}_{X} = x^{-1}$. We can also derive \eqref{eq: 1/2-BPS GGE} by summing over residues of the generating function
\begin{equation}
    \mathcal{Z}(\zeta; x) = \sum_{N=0}^\infty \zeta^N Z_N = \prod_{n=1}^\infty \frac{1}{1 - \zeta x^n}
\end{equation}
in the $\zeta$-plane.

One can intuit the fact that excitations of giant gravitons carry an opposite R-charge compared to gauge theory excitations as follows. Consider the ways of modifying the determinant operator
\begin{equation}
    \det X = \frac{1}{N!} \epsilon_{i_1 i_2 \cdots i_N} \epsilon^{j_1 j_2 \cdots j_N} X^{i_1}_{j_1} X^{i_2}_{j_2} \cdots X^{i_N}_{j_N}
\end{equation}
by replacing one of the $X$s with other fields in a large $N$ gauge theory. ``Closed-string'' modifications of the form $X \to X^{m+1}$ are trivial because they yield
\begin{equation}
    \det X \to \Tr{ X^{m} }  \det X
\end{equation}
and multi-trace operators are already accounted by the overall factor $Z_\infty$. In the half-BPS sector, the only nontrivial way of modifying $\det X$ is to replace an $X$ by the identity $I$. Such a modification effectively takes away an R-charge, so excitations $X \to I$ carry the fugacity $x^{-1}$. From a bulk point of view, a giant graviton brane $\det X$ that wraps a maximal $S^3 \subset S^5$ is saturated with $N$ units of the R-charge. It can only lose R-charges, so its open string excitations are ``holes'' in the sea of R-charges. The holes correspond to modifications that replace a single $X$ by $I$ in the determinant. \footnote{This intuition is only precise at large $N$.}

In preparation for more involved examples, it will be instructive to examine the integral definition of $\hat{Z}_k$ directly:
\begin{align} \label{eq: 1/2-BPS Zhat}
    \hat{Z}_k &= \frac{1}{k!} \oint \prod_a \frac{d\sigma_a}{2 \pi i \sigma_a} \prod_{a \neq b} (1 - \sigma_a/\sigma_b ) \prod_{a,b} \frac{1}{(1 - x^{-1} \sigma_a/\sigma_b)} \nonumber \\
    &= \frac{1}{k!} (\mathrm{diag}) \times \oint \prod_a \frac{d\sigma_a}{2 \pi i \sigma_a} \prod_{a \neq b} \frac{(\sigma_a - \sigma_b )}{(\sigma_a - x^{-1} \sigma_b)}
\end{align}
As explained in Section \ref{subsec: analytic continuation}, we analytically continue the counting of determinant modifications with our the choice of the integration cycle, which is specified through $g$. The denominators of \eqref{eq: 1/2-BPS Zhat} have physical meaning: they are open strings ending on the giant graviton $(\det X)_a$. The fugacity $x^{-1}$ conveys that the string sources negative R-charges on the giant graviton.

We can be explicit about the integration cycle by specifying denominator partitions in the theory of multivariate residues. With ordering $g = \{ g_1, \cdots, g_k \}$, the denominator partition should be chosen as
\begin{align}
    g_1 &= \sigma_1 \prod_{b \neq 1} (\sigma_1 - x^{-1} \sigma_b) \nonumber \\
    g_2 &= \sigma_2 \prod_{b \neq 2} (\sigma_2 - x^{-1} \sigma_b) \nonumber \\
    & \quad \vdots \nonumber \\
    g_k &= \sigma_k \prod_{b \neq k} (\sigma_k - x^{-1} \sigma_b),
\end{align}
We observe that any given brane $(\det X)_a$ receives a negative unit of the R-charge due to adjoint open strings that start at all other branes and end on $(\det X)_a$. There is also a zeromode $\sigma_a$ due to strings that start and end on $(\det X)_a$.

The integrand \eqref{eq: 1/2-BPS Zhat} has a degenerate pole at $\sigma_1 = \cdots = \sigma_k = 0 $ where all the desired poles condense. It is possible to compute the residue at this degenerate pole using techniques from multivariate residues.\footnote{In practice, it is more computationally efficient to resolve the degenerate pole into nondegenerate poles and sum their contributions. See Appendix \ref{app: multivariate residues}.} Following this prescription, one can indeed verify that the integral definition of $\hat{Z}_k$ agrees with \eqref{eq: half-BPS hatZ}.

\begin{figure}[t]
\centering
\includegraphics[scale=0.4]{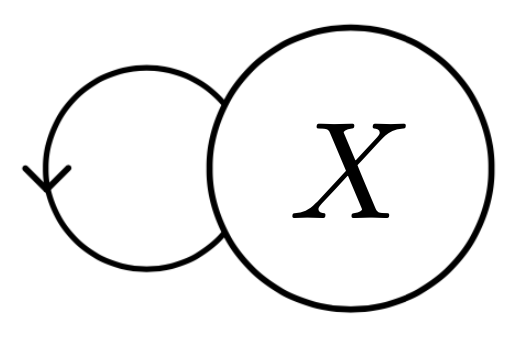}
\caption{Excitations of half-BPS giant graviton branes dual to $\det X$.}
\label{fig: half bps quiver}
\end{figure}

The exact identity \eqref{eq: 1/2-BPS GGE} suggests that the half-BPS spectrum of IIB string theory on $\AdS_5 \times S^5$ at finite $N$ is captured entirely by a sum over stacks of D3 giant gravitons dual to $(\det X)^k$, their open string excitations, and background closed strings. The bulk content is summarized in the quiver diagram in Figure \ref{fig: half bps quiver}. A property that is special to the half-BPS example is that the statement above holds at finite $\lambda$, because the half-BPS cohomology in gauge theory does not change when one turns on $\lambda$. For general BPS sectors, we expect the finite $N$ microstates in the brane expansion to survive only at the free point $\lambda = 0$.

Before moving on, let us comment on the peculiar way in which stacks of giant gravitons sum to give the half-BPS spectrum. Observe through direct power expansion that each summand in \eqref{eq: 1/2-BPS GGE} overcounts the gauge theory index $Z_N$. However due to the analytic continuation in the R-symmetry fugacity, there is an overall sign in the brane index $\hat{Z}_k$ that alternates with $k$:
\begin{equation}
    \hat{Z}_k =  \frac{(-1)^k x^{k(k+1)/2}}{\prod_{m=1}^k (1 - x^{m})}.
\end{equation}
Odd stacks of giant gravitons effectively behave as fermions in the half-BPS string theory Hilbert space, even though the half-BPS Hilbert space of the dual CFT only involves gauge-invariant combinations of bosonic scalars $X$. Vast cancellations between even and odd stacks of giant gravitons yield the finite $N$ index $Z_N$. We will discuss this further in Section \ref{subsec: brane number grading}.

\subsection{$1/4$-BPS sector} \label{subsec: quarter-bps}

Let us now consider the $1/4$-BPS index of $U(N)$ $\mathcal{N}=4$ SYM, where there is a nontrivial quiver description of microstates in the string dual. In free $\mathcal{N}=4$ SYM, the $1/4$-BPS sector is spanned by symmetrized traces of $X$ and $Y$. A way to implement the symmetrization is to count instead gauge-invariant operators made of bosonic scalars $X, Y$ carrying R-charges $(1,0)$, $(0,1)$ and a fermion $\psi$ carrying charges $(1,1)$ under $U(1)_X \times U(1)_Y \subset SO(6)_R$, equipped with the relation $Q \psi = [X, Y]$.

\begin{figure}[t]
\centering
\includegraphics[scale=0.4]{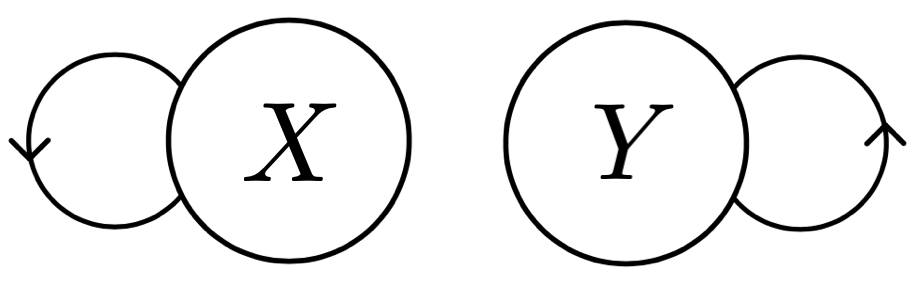}
\caption{Excitations of $1/4$-BPS giant gravitons $\det X$ and $\det Y$. It turns out that the bifundamental arrows disappear and the nodes factorize.}
\label{fig: quarter bps quiver}
\end{figure}

The adjoint single-letter index of the gauge theory in the $1/4$-BPS sector is
\begin{equation}
    f = x + y - x y.
\end{equation}
The bulk string dual is organized in terms of two orthogonal stacks of giant gravitons with the quiver diagram in Figure \ref{fig: quarter bps quiver}. Adjoint excitations of giant gravitons are given by the single-letters
\begin{align}
    \hat{f}^{X}_{X} &= \frac{x^{-1} - y}{1 - y} = \sum_{m=0}^\infty y^m (x^{-1} - y) \\
    \hat{f}^{Y}_{Y} &= \frac{y^{-1} - x}{1 - x} = \sum_{m=0}^\infty x^m (y^{-1} - x)
\end{align}
and bifundamental excitations are
\begin{align}
    \hat{f}^{X}_{Y} &= y^{-1} \\
    \hat{f}^{Y}_{X} &= x^{-1}.
\end{align}
Modifications of $\det X$ carry fugacities $x^{-1}$ and powers of $y$, so excitations of $(\det X)^k$ may take away $U(1)_X$ charges. $Y$ assumes the role of a derivative on the worldvolume theory of $(\det X)^k$. Similar statements hold for $\det Y$.

Fortunately, there is a generating function for $1/4$-BPS indices from which we can derive the giant graviton expansion \cite{Kinney:2005ej,DeSmet:2021}:
\begin{equation}
    \mathcal{Z}(\zeta; x,y) = \sum_{N=0}^\infty \zeta^N Z_N(x,y) = \frac{1}{1 - \zeta} \prod_{n=1}^\infty \frac{1}{(1 - \zeta x^n)(1 - \zeta y^n)}.
\end{equation}
Summing over residues in the $\zeta$-plane, we get
\begin{align} \label{eq: 1/4-BPS GGE}
    Z_N (x,y) &= Z_\infty \sum_{k=0}^\infty \Bigg[ x^{k N} \frac{\prod_{m = 1}^\infty (1 - y^m)}{\prod_{m=1}^k(1-x^{-m}) \prod_{m = 1}^\infty (1 - x^{-k} y^m)} \nonumber \\
    & \qquad \qquad \qquad + y^{k N} \frac{\prod_{m = 1}^\infty (1 - x^m)}{\prod_{m=1}^k (1-y^{-m}) \prod_{m = 1}^\infty (1 - y^{-k} x^m)} \Bigg] \nonumber \\
    &= Z_\infty \sum_{k=0}^\infty \left[ x^{k N} \hat{Z}_{(k,0)} + y^{k N} \hat{Z}_{(0,k)} \right]
\end{align}
where
\begin{equation}
    Z_\infty = \prod_{n=1}^\infty \frac{1}{(1-x^n)(1-y^n)}
\end{equation}
is the closed string prefactor. This expansion is convergent for $|x|,|y| < 1$. Spurious poles on the right-hand side of \eqref{eq: 1/4-BPS GGE} cancel when multiple terms are combined.

Interestingly, the $X$ and $Y$ quiver nodes factorize such that the $U(k_X) \times U(k_Y)$ mixed terms $x^{k_X N} y^{k_Y N} \hat{Z}_{(k_X,k_Y)}$ vanish. The factorization is somewhat special to the $1/4$-BPS sector at the free point of $\mathcal{N}=4$ SYM. A brane expansion of the $1/4$-BPS partition function at small $\lambda$ indeed does not factorize, as can be derived from generating functions in \cite{Kinney:2005ej}.

We can reproduce the brane indices of \eqref{eq: 1/4-BPS GGE} directly. The integral definition of the brane quiver index is
\begin{align}
    \hat{Z}_{(k_X,k_Y)} = \frac{1}{k_X! k_Y!} \oint \prod_{a=1}^{k_X} &\frac{d\sigma_{a}^{X}}{2 \pi i \sigma_{a}^{X}} \prod_{c=1}^{k_Y} \frac{d\sigma_{c}^{Y}}{2 \pi i \sigma_{c}^{Y}} \prod_{a \neq b} (1 - \sigma^{X}_a/\sigma^{X}_b) \prod_{c \neq d} (1 - \sigma^{Y}_c/\sigma^{Y}_d) \nonumber \\
    &\prod_{a,b=1}^{k_X} \prod_{m=0}^\infty \frac{1 - y^{m+1} \sigma^{X}_a / \sigma^{X}_b}{1 - x^{-1} y^{m} \sigma^{X}_a / \sigma^{X}_b} \prod_{c,d=1}^{k_Y} \prod_{m=0}^\infty \frac{1 - x^{m+1} \sigma^{Y}_c / \sigma^{Y}_d}{1 - y^{-1} x^{m} \sigma^{Y}_c / \sigma^{Y}_d} \nonumber \\
    &\prod_{a=1}^{k_X} \prod_{c=1}^{k_Y} \frac{1}{1 - y^{-1} \sigma^{X}_a / \sigma^{Y}_c} \frac{1}{1 - x^{-1} \sigma^{Y}_c / \sigma^{X}_a}.
\end{align}
Rewriting, we have
\begin{align}
    \hat{Z}_{(k_X,k_Y)} = \frac{1}{k_X! k_Y!} (\mathrm{diag}) \oint \prod_{a=1}^{k_X} &\frac{d\sigma_{a}^{X}}{2 \pi i \sigma_{a}^{X}} \prod_{c=1}^{k_Y}  \frac{d\sigma_{c}^{Y}}{2 \pi i \sigma_{c}^{Y}} \prod_{a \neq b} (1 - \sigma^{X}_a/\sigma^{X}_b) \prod_{c \neq d} (1 - \sigma^{Y}_c/\sigma^{Y}_d) \nonumber \\
    &\prod_{a \neq b} \prod_{m=0}^\infty \frac{\sigma^{X}_a - y^{m+1} \sigma^{X}_b}{\sigma^{X}_a - x^{-1} y^{m} \sigma^{X}_b} \prod_{c \neq d} \prod_{m=0}^\infty \frac{\sigma^{Y}_c - x^{m+1} \sigma^{Y}_d}{\sigma^{Y}_c - y^{-1} x^{m} \sigma^{Y}_d} \nonumber \\
    &\prod_{a=1}^{k_X} \prod_{c=1}^{k_Y} \frac{\sigma^{Y}_c}{(\sigma^{Y}_c - y^{-1} \sigma^{X}_a)} \frac{\sigma^{X}_a}{(\sigma^{X}_a - x^{-1} \sigma^{Y}_c)}.
\end{align}
The full set of excitations of giant gravitons is captured by our integration cycle prescription, but we can already learn much by interpreting the denominators of the integrand as single-string excitations of branes $(\det X)^{k_X} (\det Y)^{k_Y}$ in a large $N$ string background. Later, we will choose $g$ so that it reflects the allowed excitations ending on each brane.

Denominators of the integrand come from positive terms in brane single-letter indices. Take, for example, denominators
\begin{equation} \label{eq: 1/4-BPS adjoint denom example}
    (\sigma^{X}_1 - x^{-1} y^{m} \sigma^{X}_2) (\sigma^{X}_2 - x^{-1} y^{m} \sigma^{X}_1),
\end{equation}
that tell us about adjoint interactions of determinants
\begin{align}
    (\det X)_1 &= \frac{1}{N!} \ \epsilon_{i_1 i_2 \cdots } \epsilon^{j_1 j_2 \cdots} \ X^{i_1}_{j_1} X^{i_2}_{j_2} \cdots \nonumber \\
    (\det X)_2 &= \frac{1}{N!} \ \epsilon_{k_1 k_2 \cdots } \epsilon^{l_1 l_2 \cdots } \ X^{k_1}_{l_1} X^{k_2}_{l_2} \cdots.
\end{align}
Single modifications of $(\det X)_{1}$ and $(\det X)_{2}$ act respectively as
\begin{align}
    X^{i}_{j} &\to (Y^m)^{k}_{j} \nonumber \\
    X^{k}_{l} &\to (Y^m)^{i}_{l}
\end{align}
The brane $(\det X)_{1}$ lost a $U(1)_X$ charge but gained $m$ units of $U(1)_Y$ charge, and vice versa for $(\det X)_{2}$. The result is a pair of adjoint open strings connecting the two branes and sourcing the said charges.

It is also instructive to look at the bifundamental interactions between the following determinants:
\begin{align*}
    (\det X)_1 &= \frac{1}{N!} \ \epsilon_{i_1 i_2 \cdots}\epsilon^{j_1 j_2 \cdots} \ X^{i_1}_{j_1} X^{i_2}_{j_2} \cdots \\
    (\det Y)_2 &= \frac{1}{N!} \ \epsilon_{k_1 k_2 \cdots } \epsilon^{l_1 l_2 \cdots} \ Y^{k_1}_{l_1} Y^{k_2}_{l_2} \cdots.
\end{align*}
The relevant denominators are
\begin{equation} \label{eq: 1/4-BPS bifund denom example}
    (\sigma^{X}_1 - x^{-1} \sigma^{Y}_2) (\sigma^{Y}_2 - y^{-1} \sigma^{X}_1),
\end{equation}
which we can also write as
\begin{equation} \label{eq: 1/4-BPS bifund denom example 2}
    \propto (\sigma^{X}_1 - x^{-1} \sigma^{Y}_2) (\sigma^{X}_1 - y \sigma^{Y}_2).
\end{equation}
Expressions \eqref{eq: 1/4-BPS bifund denom example} and \eqref{eq: 1/4-BPS bifund denom example 2} suggest two very different possibilities for open strings that connect $(\det X)_1$ and $(\det Y)_2$ (modulo $X \leftrightarrow Y$). The first possibility from \eqref{eq: 1/4-BPS bifund denom example} is
\begin{align} \label{eq: 1/4-BPS bifund first possibility}
    X^{i}_{j} &\to I^{k}_{j} \nonumber \\
    Y^{k}_{l} &\to I^{i}_{l},
\end{align}
which is a pair of bifundamental strings connecting the two branes. The second possibility from \eqref{eq: 1/4-BPS bifund denom example 2} is
\begin{align} \label{eq: 1/4-BPS bifund second possibility}
    X^{i}_{j} &\to I^{k}_{j} \nonumber \\
    X^{i'}_{j'} &\to Y^{i}_{k} X^{i'}_{j'},
\end{align}
which has no bifundamental strings. In the latter scenario, a pair of bifundamental strings connecting the two branes pinch off from $(\det Y)_2$ to become a single adjoint string on $(\det X)_1$.

The ambiguity in selecting the correct excitation disappears when we recall from Section \ref{subsec: analytic continuation} that the branes are ``numbered'' via the ordering of $g$. If $g$ is ordered as
\begin{equation} \label{eq: 1/4 bps g ordering}
    g = \{ g_1^X, \cdots, g_{k_X}^X, g_1^Y, \cdots, g_{k_Y}^Y \},
\end{equation}
all allowed modifications of $(\det X)_1$ should be taken into account before considering modifications of $(\det Y)_2$. This means that the latter scenario \eqref{eq: 1/4-BPS bifund second possibility} with no bifundamental strings is the correct one. Remarkably, we were able to deduce the factorization of the $1/4$-BPS quiver in Figure \ref{fig: quarter bps quiver} just by reasoning through the allowed single-letter modifications. In general, brane quivers do not factorize because there are bifundamental excitations that are allowed only on one brane but not the other, and vice versa.

In the case of $\hat{Z}_{(1,1)}$, we can be explicit because the gauge integral reduces to an ordinary contour integral with only two poles:
\begin{equation}
    \hat{Z}_{(1,1)} = \frac{(\mathrm{diag})}{2 \pi i} \oint \frac{d\sigma_{1}^{X}}{(1 - y^{-1} \sigma^{X}_1)(\sigma^{X}_1 - x^{-1})}.
\end{equation}
We set $\sigma^{Y}_1 = 1$ in the integrand due to decoupling of one of the $U(1)$. The integrand has poles of opposite residues at $\sigma_1^X = y$ and $\sigma_1^X = x^{-1}$. In our residue prescription, the first string configuration \eqref{eq: 1/4-BPS bifund first possibility} selects a contour that encloses only the pole at $\sigma_1^X = x^{-1}$, while the second configuration \eqref{eq: 1/4-BPS bifund second possibility} selects a contour that encloses both poles. The second string configuration is the correct one according to our prescription, and indeed the two residues sum to zero. For higher mixed terms, we can only describe the integration cycle using denominator partitions.

With this intuition at hand, let us specify the integration cycle. With $g$ ordered as \eqref{eq: 1/4 bps g ordering}, the denominators should be partitioned as
{\allowdisplaybreaks
\begin{align*}
    g_1^X &= \sigma_1^X \prod_{\substack{a=1 \\ a \neq 1}}^{k_X} \prod_{m=0}^\infty (\sigma_1^X - x^{-1} y^m \sigma_a^X) \prod_{c=1}^{k_Y} (\sigma^{Y}_c - y^{-1} \sigma^{X}_1) (\sigma^{X}_1 - x^{-1} \sigma^{Y}_c) \\
    & \quad \vdots \\
    g_{k_X}^X &= \sigma_{k_X}^X \prod_{\substack{a=1 \\ a \neq k_X}}^{k_X} \prod_{m=0}^\infty (\sigma_{k_X}^X - x^{-1} y^m \sigma_a^X) \prod_{c=1}^{k_Y} (\sigma^{Y}_c - y^{-1} \sigma^{X}_{k_X}) (\sigma^{X}_{k_X} - x^{-1} \sigma^{Y}_c) \\
    g_1^Y &= \sigma_1^Y \prod_{\substack{c=1 \\ c \neq 1}}^{k_Y} \prod_{m=0}^\infty (\sigma_1^Y - y^{-1} x^m \sigma_c^Y) \\
    & \quad \vdots \\
    g_{k_Y}^Y &= \sigma_{k_Y}^Y \prod_{\substack{c=1 \\ c \neq k_Y}}^{k_Y} \prod_{m=0}^\infty (\sigma_{k_Y}^Y - y^{-1} x^m \sigma_c^Y).
\end{align*}}

\noindent Again, elements of $g$ have a common degenerate zero at $\sigma_a^X = \sigma_c^Y = 0$ where all the physical poles of the index integrand condense. One can evaluate the residue at the degenerate pole by truncating the infinite products in the denominators and working in power series. Mixed terms $\hat{Z}_{(k_X,k_Y)}$ with nonzero $k_X, k_Y$ vanish.

Working in power series, we verified that the integral definitions of $\hat{Z}_{(k,0)}, \hat{Z}_{(0,k)}$ agree with the exact answer in $\eqref{eq: 1/4-BPS GGE}$ up to $k = 3$. The agreement also holds in various ``coincident'' specializations such as $x^m = y^n$ for $m,n \in \mathbb{Z}_{\geq 0}$ with careful regularization. When fugacities are taken to be rational powers of one another, branes form bound states in the bulk. We discuss this phenomenon in Section \ref{subsec: black holes and wall-crossing}.

The convergent expansion \eqref{eq: 1/4-BPS GGE} says that the $1/4$-BPS index of IIB string theory at finite $N$ is captured entirely by sums of stacks of D3 giant gravitons dual to $(\det X)^k$ and $(\det Y)^k$ and open/closed string excitations thereof. It would be very interesting to know whether these microstates span the free $1/4$-BPS Hilbert space.

\subsection{$U(N)$-gauged fermion} \label{subsec: gauged fermion}

We now consider the index of a $U(N)$-gauged adjoint fermion $\psi$ with charge $1$ under some global symmetry $U(1)$. This theory has the single-letter index
\begin{equation}
    f = - q.
\end{equation}
Unlike in previous examples, there is no bosonic adjoint scalar with which to form a determinant operator, and fermion determinants such as $\det \psi$ vanish. However, we can still consider nonvanishing operators $\det \psi^2$ composed of fermion bilinears. These heavy fermion condensates should be dual to giant graviton-like branes in a dual string theory.

The appearance of these determinants seems quite ubiquitous in physical theories. In coming sections, we show that heavy fermion condensates play an essential role in string duals of (holomorphic twists of) $\mathcal{N}=1$ and $\mathcal{N}=2$ gauge theories.

The finite $N$ index of a $U(N)$-gauged fermion can be written exactly as
\begin{equation}
    Z_N(q) = \prod_{n=1}^N (1 - q^{2n - 1} )
\end{equation}
In \cite{Gaiotto:2021xce}, a giant graviton-like expansion for this index was found experimentally:
\begin{align}
    Z_N(q) &= Z_\infty \sum_{k=0}^\infty q^{2 k N} \frac{q^k}{\prod_{i=1}^k ( 1 - q^{2 i} ) } \\
    &= Z_\infty \sum_{k=0}^\infty q^{2 k N} \frac{(-1)^k q^{-k^2}}{\prod_{i=1}^k ( 1 - q^{-2 i} ) },
\end{align}
but an independent derivation of brane terms from determinants was not developed there. 

\begin{figure}[t]
\centering
\includegraphics[scale=0.4]{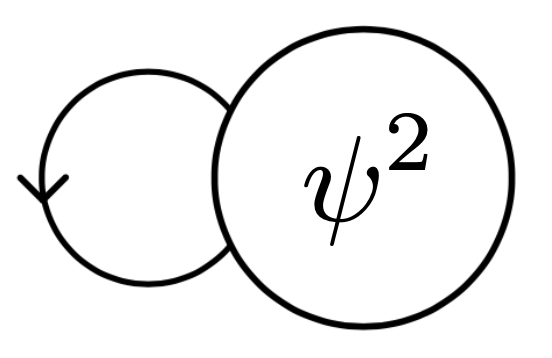}
\caption{Excitations of branes dual to a heavy fermion condensate $\det \psi^2$.}
\label{fig: fermion quiver}
\end{figure}

Here, let us derive the brane indices $\hat{Z}_k$ using the prescription in Section \ref{subsec: counting open strings}. The prescription is a simple generalization of the counting problem for bosonic determinants $\det X$ (see Appendix \ref{app: derivation half-bps}) to heavy fermion condensates $\det \psi^2$. These brane configurations have the quiver diagram illustrated in Figure \ref{fig: fermion quiver}.

The single-letter index describing modifications of $\det \psi^2$ is
\begin{equation}
    \hat{f}_{\psi^2}^{\psi^2} = q + \frac{1}{q^2} - \frac{1}{q}
\end{equation}
and the index of the gauge theory on the giant worldvolume is
\begin{align} \label{eq: fermion hatZ}
    \hat{Z}_k &= \frac{1}{k!} \oint \prod_{a=1}^k \frac{d\sigma_a}{2 \pi i \sigma_a} \prod_{a \neq b} (1 - \sigma_a /\sigma_b) \prod_{a,b=1}^k \frac{(1- q^{-1} \sigma_a / \sigma_b)}{(1 - q \sigma_a / \sigma_b)(1 - q^{-2} \sigma_a / \sigma_b)} \nonumber \\
    &= \frac{(-1)^{k}}{k!} \frac{q^{-k^2}}{(1 - q^{-2})^k} \oint \prod_{a=1}^k \frac{d\sigma_a}{2 \pi i \sigma_a} \prod_{a \neq b}  \frac{(\sigma_a - \sigma_b)}{(\sigma_a - q^{-2} \sigma_b)}.
\end{align}
$\hat{Z}_k$ comes with the prefactor $q^{2kN}$, the bare fugacity associated to $(\det \psi^2)^k$.

Notice that the integral \eqref{eq: fermion hatZ} has the same form as the half-BPS integral \eqref{eq: half-BPS hatZ}. Indeed, ``good'' modifications $\psi^2 \to I$ of the determinant possess the effective fugacity $q^{-2}$, as we are taking away two units of the $U(1)$-charge. Choosing integration contours in the same way as in the half-BPS example with $x^{-1}$ replaced by $q^{-2}$, we find
\begin{equation}
    \hat{Z}_k = \frac{(-1)^k q^{-k^2}}{\prod_{i=1}^k ( 1 - q^{-2 i} ) }.
\end{equation}
via explicit computation up to $k=3$.

Before moving on, it will be instructive to consider a similar theory with
\begin{equation}
    f = \frac{- q}{1 - q}.
\end{equation}
where now the theory of a $U(N)$-gauged fermion has derivatives with the same $U(1)$ charge as $\psi$. This gauge theory index turns out to be independent of $N$ for $N>0$. Interestingly, we can understand why there are no finite $N$ corrections by studying the brane single-letter index of $\det \psi^2$:
\begin{equation}
    \hat{f}_{\psi^2}^{\psi^2} = -1 + \frac{1}{q^2} - \frac{1}{q} + 2 q + q^2 - q^3.
\end{equation}
The brane index has an extra fermion zeromode $-1$, which in its gauge integrand becomes a vanishing numerator
\begin{equation}
    \prod_{a,b=1}^k \left( 1 - \frac{\sigma_a}{\sigma_b} \right) = 0.
\end{equation}
So the bulk reason why this $U(N)$ index does not depend on $N$ is that there are no branes that can introduce finite $N$ corrections. The brane dual of $\det \psi^2$ vanishes due to an extra fermion zeromode. In more involved examples, the presence of an extra zeromode in the adjoint single-letter index will eliminate candidates for relevant determinants.

\subsection{$\mathcal{N}=1$ vectormultiplet} \label{subsec: n1 yang mills}

We now consider the index of a four-dimensional pure $U(N)$ Yang-Mills theory consisting of a $\mathcal{N}=1$ vectormultiplet. Due to its anomalous R-symmetry, we cannot define the index of this theory on $S^1 \times S^3$. However, the vectormultiplet index we discuss can be understood as an index counting gauge-invariant local operators in the holomorphic twist of pure $\mathcal{N}=1$ Yang-Mills on $\mathbb{C}^2$ \cite{Costello:2013zra,Saberi:2019ghy}, where the R-symmetry anomaly is perturbatively Q-exact. This example displays nontrivial interactions between branes with large angular momenta dual to determinants of a field strength component and wrapped branes dual to heavy fermion condensates.

The contributing fields satisfy the BPS condition $\Delta - 2 j_1 + \frac{3}{2} r = 0$ that relates the conformal dimension $\Delta$ with global symmetry charges. Their counting is weighted by
\begin{equation}
    (-1)^{F} p^{j_1 + j_2 - \frac{1}{2} r} q^{j_1 - j_2 - \frac{1}{2} r}.
\end{equation}
There are two gauginos $\lambda_+, \lambda_-$, field strength component $F \equiv F_{++}$, and BPS derivatives $\partial_+, \partial_-$ thereof. These fields transform under Lorentz symmetry $U(1)_1 \times U(1)_2 \subset SU(2)_1 \times SU(2)_2 \simeq SO(4)$ as well as a R-symmetry $U(1)_r$ that is broken to a discrete subgroup in the quantum theory. The angular momenta and charges of the fields are summarized in Table \ref{table: pure N=1 YM fields}. 

\begin{table}[t]
\centering
\begin{tabular}{|c || c | c | c | c || c |} 
 \hline
  & $\Delta$ & $j_1$ & $j_2$ & $r$ & $f$ \\
 \hline\hline
 $\lambda_+$ & $\frac{3}{2}$ & $0$ & $\frac{1}{2}$ & $-1$ & $-p$ \\ 
 \hline
 $\lambda_-$ & $\frac{3}{2}$ & $0$ & $-\frac{1}{2}$ & $-1$ & $-q$ \\ 
 \hline
 $F$ & $2$ & $1$ & $0$ & $0$ & $p q = u$ \\ 
 \hline
 $\partial \lambda = 0$ & $\frac{5}{2}$ & $\frac{1}{2}$ & $0$ & $-1$ & $ p q$ \\ 
 \hline
 $\partial_+$ & $1$ & $\frac{1}{2}$ & $\frac{1}{2}$ & $0$ & $p$ \\
 \hline
 $\partial_-$ & $1$ & $\frac{1}{2}$ & $-\frac{1}{2}$ & $0$ & $q$ \\
 \hline
\end{tabular}
\caption{Letters and their charges in the $\mathcal{N}=1$ vectormultiplet index.}
\label{table: pure N=1 YM fields}
\end{table}

The field strength $F$ and gaugino equations of motion have the same fugacities, so we resolved their contributions by assigning the fugacity $u$ for $F$. The condition $u = p q$ should be imposed in the final result. Doing so resolves a divergence in the brane index that ends up cancelling out in the full sum over brane configurations. In total, the single-letter index of the $\mathcal{N}=1$ vectormultiplet is
\begin{equation}
    f = 1 - \frac{(1-u)}{(1-p)(1-q)} = \frac{- p - q + p q + u}{(1-p)(1-q)}.
\end{equation}

\begin{figure}[t]
\centering
\includegraphics[scale=0.4]{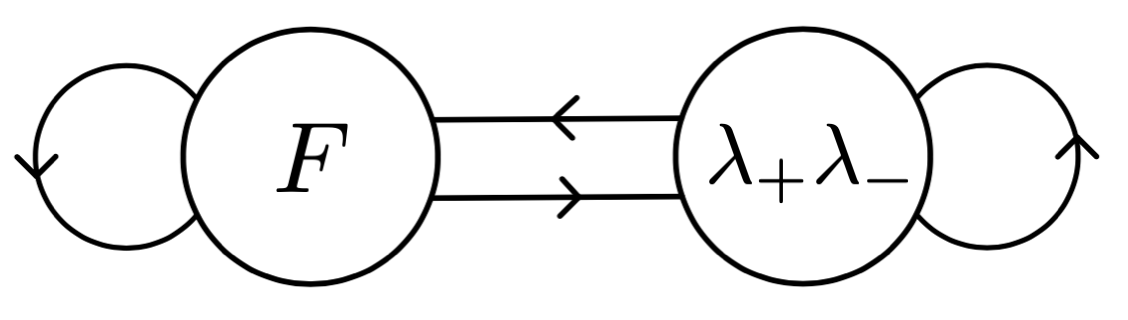}
\caption{Brane quiver for twisted pure $\mathcal{N}=1$ Yang-Mills. The branes are dual to gauge theory determinants $\det F_{++}$ and $\det \lambda_+ \lambda_-$.}
\label{fig: n1 quiver}
\end{figure}

Given the fields, let us guess a few candidates for relevant determinant operators:
\begin{itemize}
    \item $\det F$, with fugacity $u^N$
    \item $\det \lambda_+ \lambda_-$, with fugacity $p^N q^N$
    \item $\det \lambda_+^2$, with fugacity $p^{2N}$
    \item $\det \lambda_-^2$, with fugacity $q^{2N}$.
\end{itemize}
We apply the prescription of Section \ref{subsec: counting open strings} to find single-letter indices for these determinants. The adjoint indices of $\det \lambda_+^2$ and $\det \lambda_-^2$ suffer from extra fermion zeromodes $-1$, so their nodes in the quiver vanish. Adjoint indices of $\det F$ and $\det \lambda_+ \lambda_-$ survive and are
\begin{align}
    \hat{f}_{F}^{F} &= \frac{1}{u} + p + q - p q - \frac{p}{u} - \frac{q}{u} +  \frac{p q}{u} \\
    \hat{f}_{\lambda}^{\lambda} &= \frac{1}{1 - u} \left( \frac{1}{p q} + 2 p + 2 q - \frac{1}{p} - \frac{1}{q} - u - p q - p^2 q - p q^2 + p^2 q^2  \right).
\end{align}
Their bifundamental letters are
\begin{align}
    \hat{f}^{F}_{\lambda} &= \frac{1}{p q} + p + q - \frac{1}{p} - \frac{1}{q} - p q \\
    \hat{f}^{\lambda}_{F} &= \frac{1}{u} - \frac{p}{u} - \frac{q}{u} + \frac{p^2 q}{u} + \frac{p q^2}{u} - \frac{p^2 q^2}{u}.
\end{align}
Brane configurations in the string dual of twisted pure $\mathcal{N}=1$ Yang-Mills on $\mathbb{C}^2$ are given by the $U(k_{F}) \times U(k_{\lambda})$ quiver diagram of Figure \ref{fig: n1 quiver}.

For the $\mathcal{N}=1$ vectormultiplet, we find the brane expansion
\begin{equation} \label{eq: N=1 vector GGE}
    Z_N = Z_\infty \sum_{k_F,k_\lambda = 0}^\infty u^{k_F N} (p q)^{k_\lambda N} \hat{Z}_{(k_F,k_\lambda)},
\end{equation}
where $\hat{Z}_{(k_F,k_\lambda)}$ contains all open string excitations and the sum is over all ranks of the quiver gauge group. The background closed string spectrum is
\begin{equation}
    Z_\infty = \prod_{n=1}^\infty \frac{(1 - p^n)(1 - q^n)}{(1 - u^n)}.
\end{equation}
We impose the constraint $u = p q$ when evaluating the final expression.

It is straightforward to translate the single-letter indices into the full index:
{\allowdisplaybreaks
\begin{align}
    \hat{Z}_{(k_F,k_{\lambda})} = &\frac{1}{k_F! k_{\lambda}!} (\mathrm{diag}) \times \oint \prod_{a=1}^{k_F} \frac{d\sigma_{a}^{F}}{2 \pi i \sigma_{a}^{F}} \prod_{c=1}^{k_{\lambda}} \frac{d\sigma_{c}^{\lambda}}{2 \pi i \sigma_{c}^{\lambda}} \prod_{a \neq b}^{k_F} (1 - \sigma^{F}_a/\sigma^{F}_b) \prod_{c \neq d}^{k_{\lambda}} (1 - \sigma^{\lambda}_c/\sigma^{\lambda}_d) \nonumber \\
    &\prod_{a \neq b}^{k_F} \Bigg[ \frac{ \sigma_a^F \left(\sigma_a^F - p q \sigma_b^F \right) \left(\sigma_a^F - \frac{p}{u} \sigma_b^F\right) \left(\sigma_a^F - \frac{q}{u} \sigma_b^F\right)}{\left(\sigma_a^F - \frac{1}{u} \sigma_b^F\right) \left(\sigma_a^F - p \sigma_b^F\right) \left(\sigma_a^F - q \sigma_b^F\right) \left(\sigma_a^F - \frac{p q}{u} \sigma_b^F \right)} \Bigg] \nonumber \\
    &\prod_{c \neq d}^{k_{\lambda}} \prod_{m=0}^\infty \Bigg[ \frac{\left( \sigma_c^{\lambda} - \frac{u^m}{p} \sigma_d^{\lambda} \right) \left( \sigma_c^{\lambda} - \frac{u^m}{q} \sigma_d^{\lambda} \right) \left( \sigma_c^{\lambda} - u^{m+1} \sigma_d^{\lambda} \right)}{\left( \sigma_c^{\lambda} - u^m p \sigma_d^{\lambda} \right)^2 \left( \sigma_c^{\lambda} - u^m q \sigma_d^{\lambda} \right)^2} \nonumber \\
    & \qquad \qquad \frac{\left( \sigma_c^{\lambda} - u^m p q  \sigma_d^{\lambda} \right) \left( \sigma_c^{\lambda} - u^m p^2 q  \sigma_d^{\lambda} \right) \left( \sigma_c^{\lambda} - u^m p q^2  \sigma_d^{\lambda} \right)}{\left( \sigma_c^{\lambda} - \frac{u^m}{pq} \sigma_d^{\lambda} \right) \left( \sigma_c^{\lambda} - u^m p^2 q^2 \sigma_d^{\lambda} \right) } \Bigg] \nonumber \\
    &\prod_{a=1}^{k_F} \prod_{c=1}^{k_{\lambda}} \Bigg[ \frac{\left( \sigma_c^{\lambda} - p q \sigma_a^F \right) \left( \sigma_c^{\lambda} - \frac{1}{p} \sigma_a^F \right) \left( \sigma_c^{\lambda} - \frac{1}{q} \sigma_a^F \right)}{\left( \sigma_c^{\lambda} - \frac{1}{p q} \sigma_a^F \right) \left( \sigma_c^{\lambda} - p \sigma_a^F \right) \left( \sigma_c^{\lambda} - q \sigma_a^F \right)} \nonumber \\
    & \qquad \qquad \frac{\left( \sigma_a^F - \frac{p}{u} \sigma_c^{\lambda} \right) \left( \sigma_a^F - \frac{q}{u} \sigma_c^{\lambda} \right) \left( \sigma_a^F - \frac{p^2 q^2}{u} \sigma_c^{\lambda} \right)}{\left( \sigma_a^F - \frac{1}{u} \sigma_c^{\lambda} \right) \left( \sigma_a^F - \frac{p^2 q}{u} \sigma_c^{\lambda} \right) \left( \sigma_a^F - \frac{p q^2}{u} \sigma_c^{\lambda} \right)} \Bigg].
\end{align}}

\noindent The first and second brackets are adjoint terms and the third bracket contains bifundamental terms. Adjoint terms indicate that the individual brane indices have divergences at $u = p q$. These divergences will cancel in the brane expansion \eqref{eq: N=1 vector GGE} when we combine all terms $\hat{Z}_{(k_F, k_{\lambda})}$ that contribute at a given order in power series.

We find the possible open string excitations by examining the denominators. Let us start with adjoint excitations of $\det F$, for which the relevant denominators are
\begin{equation} \label{eq: N=1 vector u adjoint denom}
    \left(\sigma_a^F - \frac{1}{u} \sigma_b^F\right) \left(\sigma_a^F - p \sigma_b^F\right) \left(\sigma_a^F - q \sigma_b^F\right) \left(\sigma_a^F - \frac{p q}{u} \sigma_b^F \right).
\end{equation}
In the order shown, they correspond to the modifications
\begin{align} \label{eq: N=1 vector f modifications}
    F &\to I \nonumber \\
    F &\to \partial_+ F \nonumber \\
    F &\to \partial_- F \nonumber \\
    F &\to \lambda_+ \lambda_-,
\end{align}
where we suppressed adjoint matrix indices. Adjoint open strings on the brane dual of $\det F$ source charges that can be read off from \eqref{eq: N=1 vector f modifications} and Table \ref{table: pure N=1 YM fields}. Note that fugacities in the following rearrangement of \eqref{eq: N=1 vector u adjoint denom}
\begin{equation}
    \propto \left(\sigma_b^F - u \sigma_a^F\right) \left(\sigma_b^F - \frac{1}{p} \sigma_a^F\right) \left(\sigma_b^F - \frac{1}{q} \sigma_a^F\right) \left(\sigma_b^F - \frac{u}{p q} \sigma_a^F \right).
\end{equation}
do not correspond to any allowed modification of $\det F$.

For adjoint excitations of $\det \lambda_+ \lambda_-$ we have denominators
\begin{equation} \label{eq: N=1 vector pq adjoint denom}
    \left( \sigma_c^{\lambda} - u^m p \sigma_d^{\lambda} \right)^2 \left( \sigma_c^{\lambda} - u^m q \sigma_d^{\lambda} \right)^2 \left( \sigma_c^{\lambda} - \frac{u^m}{p q} \sigma_d^{\lambda} \right) \left( \sigma_c^{\lambda} - u^m p^2 q^2 \sigma_d^{\lambda} \right),
\end{equation}
which correspond to modifications
\begin{align} \label{eq: N=1 vector gaugino modifications}
    \lambda_+ \lambda_- &\to F^m (\partial_+ \lambda_+) \lambda_- \nonumber \\
    \lambda_+ \lambda_- &\to F^m \lambda_+ (\partial_- \lambda_-) \nonumber \\
    \lambda_+ \lambda_- &\to F^m \nonumber \\
    \lambda_+ \lambda_- &\to F^m (\partial_+^2 \lambda_+)(\partial_-^2 \lambda_-)
\end{align}
modulo double poles.\footnote{Modifications cannot be proportional to the field being replaced, because the extra letters would factor out from the determinant as a trace and traces are already accounted by the closed string spectrum $Z_\infty$. Also, the derivatives were taken to act on gauginos in a way that no ambiguities arise due to gaugino equations of motion.} $F$ assumes the role of a derivative in the $U(k_{\lambda})$ worldvolume theory on branes dual to $(\det \lambda_+ \lambda_-)^k$. It is again clear that inverses of fugacities in \eqref{eq: N=1 vector pq adjoint denom} do not correspond to allowed excitations. 

We can also find bifundamental excitations from the (rearranged) denominators
\begin{equation}
    \propto \left( \sigma_a^F - \frac{1}{u} \sigma_c^{\lambda} \right) \left( \sigma_a^F - \frac{p^2 q}{u} \sigma_c^{\lambda} \right) \left( \sigma_a^F - \frac{p q^2}{u} \sigma_c^{\lambda} \right) \left( \sigma_a^F - p q \sigma_c^{\lambda} \right) \left( \sigma_c^{\lambda} - p \sigma_a^F \right) \left( \sigma_c^{\lambda} - q \sigma_a^F \right).
\end{equation}
In order, they correspond to
\begin{align} \label{eq: N=1 vector bifund mods}
    F &\to I \nonumber \\
    F &\to (\partial_+ \lambda_+) \lambda_- \nonumber \\
    F &\to \lambda_+ (\partial_- \lambda_-) \nonumber \\
    F &\to \partial_+ \partial_- F \nonumber \\
    \lambda_+ \lambda_- &\to (\partial_+ \lambda_+) \lambda_- \nonumber \\
    \lambda_+ \lambda_- &\to \lambda_+ (\partial_- \lambda_-).
\end{align}
As in the $1/4$-BPS example, bifundamental terms have some ambiguity that is resolved in our residue prescription. In particular, the modifications $F \to I$ and $F \to \partial_+ \partial_- F$ of $\det F$ could have been interpreted from the perspective of $\det \lambda_+ \lambda_-$ as
\begin{align}
    \lambda_+ \lambda_- &\to \partial_+ \partial_- F \nonumber \\
    \lambda_+ \lambda_- &\to I.
\end{align}
If $g$ is ordered as
\begin{equation}
    g = \{ g_1^F, \cdots , g_{k_F}^F, g_1^{\lambda}, \cdots, g_{k_{\lambda}}^{\lambda} \},
\end{equation}
then we should categorize the modifications of $\det F$ first before considering $\det \lambda_+ \lambda_-$. Therefore, the correct set of modifications are those in \eqref{eq: N=1 vector bifund mods}. Unlike in the $1/4$-BPS example, there are bifundamental excitations that are allowed only on $\det F$ but not on $\det \lambda_+ \lambda_-$, and vice versa. Hence, the brane quiver for the $\mathcal{N}=1$ vectormultiplet does not factorize.

Let us summarize the allowed excitations by writing the denominator partition $g$:
\begin{align}
    g_a^F &= \sigma_a^F \prod_{\substack{b=1 \\ b \neq a}}^{k_F} \left(\sigma_a^F - \frac{1}{u} \sigma_b^F\right) \left(\sigma_a^F - p \sigma_b^F\right) \left(\sigma_a^F - q \sigma_b^F\right) \left(\sigma_a^F - \frac{p q}{u} \sigma_b^F \right) \nonumber \\
    & \qquad \quad \prod_{c=1}^{k_{\lambda}} \left( \sigma_a^F - \frac{1}{u} \sigma_c^{\lambda} \right) \left( \sigma_a^F - \frac{p^2 q}{u} \sigma_c^{\lambda} \right) \left( \sigma_a^F - \frac{p q^2}{u} \sigma_c^{\lambda} \right) \left( \sigma_c^{\lambda} - \frac{1}{p q} \sigma_a^F \right) \nonumber \\
    g_c^{\lambda} &= \sigma_c^{\lambda} \prod_{\substack{d=1 \\ d \neq c}}^{k_{\lambda}} \prod_{m=0}^\infty \left( \sigma_c^{\lambda} - u^m p \sigma_d^{\lambda} \right)^2 \left( \sigma_c^{\lambda} - u^m q \sigma_d^{\lambda} \right)^2 \left( \sigma_c^{\lambda} - \frac{u^m}{pq} \sigma_d^{\lambda} \right) \left( \sigma_c^{\lambda} - u^m p^2 q^2 \sigma_d^{\lambda} \right) \nonumber \\
    & \qquad \quad \prod_{a=1}^{k_F} \left( \sigma_c^{\lambda} - p \sigma_a^F \right) \left( \sigma_c^{\lambda} - q \sigma_a^F \right),
\end{align}
where $a,b = 1, 2, \cdots, k_F$ and $c,d = 1, 2, \cdots, k_{\lambda}$. The denominator partition $g$ uniquely specifies the integration cycle around the degenerate pole $\sigma_a^F = \sigma_c^{\lambda} = 0$. In Appendix \ref{app: n1 yang mills checks}, we present checks of the brane expansion \eqref{eq: N=1 vector GGE} by evaluating the brane indices in power series.

So far, our gauge theory prescription gave us indices of branes in a tentative string dual of a twisted pure $\mathcal{N}=1$ vectormultiplet. Explicit computations of the brane expansion shows that the sum over brane duals of $\det F$ and $\det \lambda_+ \lambda_-$ (and their excitations) are sufficient to explain the index spectrum of this string theory, up to the order checked in Appendix \ref{app: n1 yang mills checks}. We emphasize that we did not need any knowledge of the string dual thus far. It is nevertheless interesting to interpret the determinants in a known string background. We should mention that the picture in terms of wrapped branes in a background geometry is only valid at large $N$. On the other hand, brane indices are well-defined at finite $N$ and are in fact independent of $N$ apart from its bare fugacity prefactor. 

String backgrounds contain cycles whose isometries are generated by global symmetry charges of the gauge theory. The branes we are interested in are analogous to maximal giant gravitons. In the presence of RR field strengths, these branes acquire an angular momentum of order $N$ at fixed points of isometries \cite{McGreevy:2000cw}. For $S^n$, a brane wrapped on the fixed $S^{n-2}$ sphere of a $U(1)$ isometry maximizes the charge under $U(1)$.

It was proposed in \cite{Maldacena:2000yy} (see also \cite{Klebanov:2000hb}) that pure $SU(N)$ $\mathcal{N}=1$ Yang-Mills theory is dual to the IR limit of IIB string theory with $N$ D5 branes wrapped on $S^2$ in a background that is topologically
\begin{equation*}
    \mathbb{R}^{4} \times \mathbb{R}_{\geq 0} \times S^2 \times S^3
\end{equation*}
with appropriate twists that preserve four supersymmetries. There are $N$ units of RR three-form flux across $S^3$. The wrapped $S^2$ component shrinks in the IR and blows up in the UV, while $S^3$ maintains finite radius. $S^2$ is nontrivially fibered over $S^3$ in a way that nontrivial two-cycles are extended over both $S^2$ and $S^2 \subset S^3$ in a correlated manner. Rotations of $S_{1}^1 \times S_{2}^1 \subset \tilde{S}^3 \subset \mathbb{R}^{4}$ along two orthogonal planes of $\mathbb{R}^{4}$ are isometries generated by Lorentz angular momenta $j_1, j_2$. Rotation along a certain axis of $S^3$ corresponds to the action of r-symmetry $U(1)_r$, which is broken to $\mathbb{Z}_{2N}$ in the UV by D1 instantons wrapping the nontrivial two-cycle.

The heavy gaugino condensate $\det \lambda_+ \lambda_-$ starts out with $\sim N$ units of r-charge and no angular momenta $j_1,j_2$ (see Table \ref{table: pure N=1 YM determinants}). Its excitations \eqref{eq: N=1 vector gaugino modifications} and \eqref{eq: N=1 vector bifund mods} suggest that it can gain an arbitrary amount of angular momenta and that it can lose but not gain r-charges. Therefore, the brane dual of $\det \lambda_+ \lambda_-$ sits in the time direction $\mathbb{R}_t$ and wraps a maximal circle $S^1$ in $S^3$ that is fixed under the $U(1)_r$ isometry. The bound on the r-charge reflects the finite size of this circle. As it gains Lorentz angular momentum, its trajectory can rotate in the spatial planes along the circle $S_1^1$, because the worldvolume derivatives $F^m$ supply $j_1$ charges. A natural candidate for the brane dual is a D1 brane that wraps $\mathbb{R}_t \times S^1$, where $S^1$ is the fixed circle in $S^3$ under $U(1)_r$.

The field strength determinant $\det F$ starts out with $N$ units of angular momentum $j_1$. Its excitations \eqref{eq: N=1 vector f modifications} and \eqref{eq: N=1 vector bifund mods} suggest that it can gain r-charges $r$ and angular momentum $j_2$, and that it can gain or lose $j_1$. The brane dual of $\det F$ thus sits in the time direction $\mathbb{R}_t$ and wraps a large spatial circle in $\tilde{S}^3$ that is fixed under $U(1)_{1}$. The radius of the wrapped circle changes as the brane gains or loses angular momentum $j_1$. Angular momenta and size of the brane can grow without bound. Natural candidates for the brane dual are D1 branes wrapping $\mathbb{R}_t \times S^1$, where $S^1$ is a circle in $\tilde{S}^3$ fixed under $U(1)_1$, or D3 branes that also wrap the nontrivial two-cycle.

\begin{table}[t]
\centering
\begin{tabular}{|c || c | c | c | c || c |} 
 \hline
  & $\Delta$ & $j_1$ & $j_2$ & $r$ & $\mathrm{fugacity}$ \\
 \hline\hline
 $\det \lambda_+ \lambda_-$ & $3N$ & $0$ & $0$ & $-2 N$ & $p^N q^N$ \\ 
 \hline
 $\det F$ & $2 N$ & $N$ & $0$ & $0$ & $p^N q^N = u^N$ \\ 
 \hline
\end{tabular}
\caption{Charges of determinant operators in the $\mathcal{N}=1$ vectormultiplet.}
\label{table: pure N=1 YM determinants}
\end{table}

\subsection{$\mathcal{N}=2$ vectormultiplet} \label{subsec: n2 yang mills}

Let us now consider the index of pure $U(N)$ $\mathcal{N}=2$ Yang-Mills theory consisting of a single $\mathcal{N}=2$ vectormultiplet. Again due to the anomalous R-symmetry, the $\mathcal{N}=2$ vectormultiplet index can be understood as that counting gauge-invariant local operators in the holomorphic twist of pure $\mathcal{N}=2$ Yang-Mills on $\mathbb{C}^2$ \cite{Costello:2013zra,Saberi:2019ghy}. This example will involve several nontrivial heavy fermion condensates such as those formed from a gaugino-chiralino pair.

The contributing fields satisfy the BPS condition $\Delta - 2 j_1 - 2 R + \frac{1}{2} r = 0$ that relates the conformal dimension $\Delta$ with global symmetry charges. Their counting is weighted as
\begin{equation}
    (-1)^{F} p^{j_1 + j_2 - \frac{1}{2} r} q^{j_1 - j_2 - \frac{1}{2} r} t^{R + \frac{1}{2}r}.
\end{equation}
There is a $N=1$ chiral multiplet scalar $X$ and its partner fermion $\psi$, as well as fields and derivatives seen in the $\mathcal{N}=1$ vectormultiplet. The fields are charged under Lorentz symmetry $U(1)_1 \times U(1)_2 \subset SU(2)_1 \times SU(2)_2 \simeq SO(4)$ as well as R-symmetries $SU(2)_R \times U(1)_r$. $U(1)_r$ is again broken to a discrete subgroup $\mathbb{Z}_{4N}$ in the quantum theory. Fields and their charges are summarized in Table \ref{table: pure N=2 YM fields}.

The single-letter index of a $\mathcal{N}=2$ vectormultiplet is
\begin{equation}
    f = 1 - \frac{(1-x)(1+t)}{(1-p)(1-q)} = \frac{-p - q + p q + x t + x - t}{(1-p)(1-q)}
\end{equation}
with the constraint $x t = p q$. As before, we resolved equal-fugacity contributions to the index from $F$ and the gaugino equations of motion.\footnote{In a standard basis of charges considered in the literature, the $\mathcal{N}=2$ vectormultiplet index is
\begin{equation}
    f = \frac{-p - q + 2 p q + (p q)^{1/3} \mathsf{x}^{-1} - (p q)^{2/3} \mathsf{x}}{(1-p)(1-q)}.
\end{equation}
The fugacities there are related to our fugacities as $x = (p q)^{1/3} \mathsf{x}^{-1}$ and $t = (p q)^{2/3} \mathsf{x}$, where we further impose $x t = p q$.} The resolution regulates intermediate divergences in brane indices that cancel out in the end.

\begin{table}[t]
\centering
\begin{tabular}{|c || c | c | c | c | c || c |} 
 \hline
  & $\Delta$ & $j_1$ & $j_2$ & $R$ & $r$ & $f$ \\
 \hline\hline
 $X$ & $1$ & $0$ & $0$ & $0$ & $-2$ & $p q/t = x$ \\ 
 \hline
 $\psi$ & $\frac{3}{2}$ & $\frac{1}{2}$ & $0$ & $\frac{1}{2}$ & $1$ & $-t$ \\ 
 \hline
 $\lambda_+$ & $\frac{3}{2}$ & $0$ & $\frac{1}{2}$ & $\frac{1}{2}$ & $-1$ & $-p$ \\ 
 \hline
 $\lambda_-$ & $\frac{3}{2}$ & $0$ & $-\frac{1}{2}$ & $\frac{1}{2}$ & $-1$ & $-q$ \\ 
 \hline
 $F$ & $2$ & $1$ & $0$ & $0$ & $0$ & $p q = x t$ \\ 
 \hline
 $\partial \lambda = 0$ & $\frac{5}{2}$ & $\frac{1}{2}$ & $0$ & $\frac{1}{2}$ & $-1$ & $ p q$ \\ 
 \hline
 $\partial_+$ & $1$ & $\frac{1}{2}$ & $\frac{1}{2}$ & $0$ & $0$ & $p$ \\
 \hline
 $\partial_-$ & $1$ & $\frac{1}{2}$ & $-\frac{1}{2}$ & $0$ & $0$ & $q$ \\
 \hline
\end{tabular}
\caption{Letters and their charges in the $\mathcal{N}=2$ vectormultiplet index.}
\label{table: pure N=2 YM fields}
\end{table}

Branes whose adjoint indices do not have extra zeromodes correspond to the following determinant operators in gauge theory:
\begin{itemize}
    \item $\det X$, with fugacity $x^N$
    \item $\det \psi^2$, with fugacity $t^{2N}$
    \item $\det \lambda_+ \lambda_-$, with fugacity $p^N q^N$
    \item $\det \lambda_+ \psi$, with fugacity $p^N t^N$
    \item $\det \lambda_- \psi$, with fugacity $q^N t^N$.
\end{itemize}
It is simple to work out the single-letter indices as well as the full brane index
\begin{equation}
    \hat{Z}_{( k_{X}, k_{\psi^2},  k_{\lambda_+ \lambda_-},  k_{\lambda_+ \psi},  k_{\lambda_- \psi} )}.
\end{equation}
Intriguingly, we find that all five nodes of the quiver given by the determinants above are necessary to reproduce the finite $N$ gauge theory index. We provide checks of our proposal using the brane expansion in Appendix \ref{app: n2 yang mills checks}, up to the second set of corrections.

For illustration, we present the integration cycles only for $\hat{Z}_{(k,0,0,0,0)}$. It is straightforward but tedious to work out the cycles in the general case. The single-letter index for adjoint excitations of $\det X$ is
\begin{equation}
    \hat{f}_X^X = \frac{1}{1+t} \left( p + q - p q + t + \frac{1}{x} - \frac{p}{x} - \frac{q}{x} + \frac{pq}{x} \right)
\end{equation}
A peculiarity in this example is that expanding $1/(1+t)$ will result in alternating signs. The signs invert the infinite product in the gauge integrand in an alternating fashion. The physical meaning of this feature at the level of the worldvolume gauge theory will become clear when we write the explicit modifications.

The index of $(\det X)^k$ is
\begin{align}
    \hat{Z}_{(k,0,0,0,0)} = &\frac{1}{k!} \mathrm{(diag)} \times \oint \prod_{a=1}^k \frac{d\sigma_a}{2 \pi i \sigma_a} \prod_{a \neq b} \left( 1 - \sigma_a \sigma_b^{-1} \right) \nonumber \\
    &\prod_{a \neq b}^k \prod_{m=0}^\infty \Bigg[ \frac{(\sigma_a - p q t^{2 m} \sigma_b) (\sigma_a - \frac{p}{x} t^{2 m} \sigma_b) (\sigma_a - \frac{q}{x} t^{2 m} \sigma_b) }{ (\sigma_a - p t^{2 m} \sigma_b) (\sigma_a - q t^{2 m} \sigma_b) (\sigma_a - t^{2 m+1} \sigma_b) (\sigma_a - \frac{1}{x} t^{2 m} \sigma_b) (\sigma_a - \frac{p q}{x} t^{2 m} \sigma_b) } \nonumber \\
    & \quad \frac{ (\sigma_a - p t^{2 m+1} \sigma_b) (\sigma_a - q t^{2 m+1} \sigma_b) (\sigma_a - t^{2 m+2} \sigma_b) (\sigma_a - \frac{1}{x} t^{2 m+1} \sigma_b) (\sigma_a - \frac{p q}{x} t^{2 m+1} \sigma_b) }{(\sigma_a - p q t^{2 m+1} \sigma_b) (\sigma_a - \frac{p}{x} t^{2 m+1} \sigma_b) (\sigma_a - \frac{q}{x} t^{2 m+1} \sigma_b) } \Bigg].
\end{align}
In the order shown, denominators indicate that the modifications are\footnote{Purely based on fugacities that we considered, there can be ambiguities in identifying fugacities with physical fields. An example is the denominator $(\sigma_a - p q t^{2 m+1} \sigma_b)$ that admits several candidates. We resolved this ambiguity by hand by further refining the fugacities corresponding to fields.}
\begin{align}
    X &\to \psi^{2m} \partial_+ X \nonumber \\
    X &\to \psi^{2m} \partial_- X \nonumber \\
    X &\to \psi^{2m} F \nonumber \\
    X &\to \psi^{2m} \nonumber \\
    X &\to \psi^{2m} \lambda_+ \lambda_- \nonumber \\
    X &\to \psi^{2m} F \lambda_+ \lambda_- \nonumber \\
    X &\to \psi^{2m+1} \lambda_+ \nonumber \\
    X &\to \psi^{2m+1} \lambda_-.
\end{align}
Even powers of chiralinos $\psi$ are acting as derivatives in the worldvolume theory on branes dual to $(\det X)^k$. Large powers of adjoint fermions may seem concerning, but recall that we assume large $N$ during the derivation of the brane indices. $\psi^2$ can behave just as bosonic derivatives in the $U(k)$ worldvolume theory sitting in a large $N$ string background.

Using the theory of multivariate residues, the denominator partition $g = \{ g_1, \cdots, g_k \}$ now uniquely specifies the integration cycle for $\hat{Z}_{( k,0,0,0,0)}$:
\begin{align*}
    g_a = \sigma_a^2 \prod_{\substack{b = 1 \\ b \neq a}}^k \prod_{m=0}^\infty &(\sigma_a - p t^{2 m} \sigma_b) (\sigma_a - q t^{2 m} \sigma_b) (\sigma_a - t^{2 m+1} \sigma_b) (\sigma_a - \frac{1}{x} t^{2 m} \sigma_b) \\ &(\sigma_a - \frac{p q}{x} t^{2 m} \sigma_b) (\sigma_a - p q t^{2 m+1} \sigma_b) (\sigma_a - \frac{p}{x} t^{2 m+1} \sigma_b) (\sigma_a - \frac{q}{x} t^{2 m+1} \sigma_b) 
\end{align*}
for $a= 1, \cdots, k$. All physical poles condense at $\sigma = ( 0, \cdots, 0)$ so there is a degenerate residue. Degenerate poles can be treated directly in the theory of multivariate residues, but we resolved the poles in practical computations.

Pure $SU(N)$ $\mathcal{N}=2$ Yang-Mills theory has a string dual in terms of $N$ D5 branes wrapped on $\mathbb{R}^4 \times S^2$ \cite{Gauntlett:2001ps,Bigazzi:2001aj} in a background that is topologically
\begin{equation*}
    \mathbb{R}^4 \times \mathbb{R}_{\geq 0} \times S^2 \times S^3.
\end{equation*}
There are $N$ units of RR three-form flux across $S^3$. A difference from the $\mathcal{N}=1$ construction is that the wrapped two-cycle $S^2$ is chosen such that it preserves eight supersymmetries. More precisely, the wrapped $S^2$ is embedded in a Calabi-Yau twofold rather than a threefold. The choice of $S^2$ implements the appropriate topological twist for the $\mathcal{N}=2$ theory. Again, there are nontrivial two- and three-cycles inside $S^2 \times S^3$. Cycles generated by isometries are $S_{1}^1 \times S_{2}^1 \subset \tilde{S}^3 \subset \mathbb{R}^4$ for Lorentz rotations $j_1,j_2$ and $S_R^1 \times S_r^1 \subset S^3$ for R-symmetries $R,r$. The full $SU(2)_R$ is not visible at the level of the metric of the supergravity background, but its $U(1)_R$ subgroup is generated by translations of an angle $\phi_1$ of $S^3$ that also serves as an Euler angle rotating the $S^2$.

The nontrivial topology of the compact directions makes it difficult to comment on precise brane duals of determinant operators. However, it may be possible to find the brane duals at large $N$ based on the charges they possess (see Table \ref{table: pure N=2 YM determinants}) and from their excitations via explicit gauge theory letters.

\begin{table}[t]
\centering
\begin{tabular}{|c || c | c | c | c | c || c |} 
 \hline
  & $\Delta$ & $j_1$ & $j_2$ & $R$ & $r$ & $\mathrm{fugacity}$ \\
 \hline\hline
 $\det X$ & $N$ & $0$ & $0$ & $0$ & $-2N$ & $p^N q^N/ t^N = x^N$ \\ 
 \hline
 $\det \psi^2$ & $3N$ & $N$ & $0$ & $N$ & $2 N$ & $t^{2N}$ \\ 
 \hline
 $\det \lambda_+ \lambda_-$ & $3N$ & $0$ & $0$ & $N$ & $-2N$ & $p^N q^N$ \\ 
 \hline
 $\det \lambda_+ \psi$ & $3 N$ & $\frac{1}{2}N$ & $\frac{1}{2}N$ & $N$ & $0$ & $p^N t^N$ \\ 
 \hline
 $\det \lambda_- \psi$ & $3 N$ & $\frac{1}{2}N$ & $-\frac{1}{2}N$ & $N$ & $0$ & $q^N t^N$ \\ 
 \hline
\end{tabular}
\caption{Charges of determinant operators in the $\mathcal{N}=2$ vectormultiplet.}
\label{table: pure N=2 YM determinants}
\end{table}

\subsection{A puzzle for the $1/16$-BPS sector of $\mathcal{N}=4$ SYM} \label{subsec: 1/16 bps sector}

To conclude the list of examples, we point out a puzzle for the $1/16$-BPS index of $\mathcal{N}=4$ super Yang-Mills.

In a basis of charges, the $1/16$-BPS index is
\begin{equation}
    Z_N = \Tr_{\mathcal{H}_N} (-1)^F p^{\Delta - j_1 + j_2 - Y} q^{\Delta - j_1 - j_2 - Y} x^{-q_2 - q_3 + Y} y^{-q_3 - q_1 + Y} z^{-q_1 - q_2 + Y}
\end{equation}
constrained by $x y z = p q$. The BPS condition imposes
\begin{equation}
    \Delta - 2 j_1 - q_1 - q_2 - q_3 = 0.
\end{equation}
This basis makes manifest the angular momenta and R-charges of the fields in Table \ref{table: N=4 YM fields}. The single-letter index of the theory is
\begin{equation}
    f = 1 - \frac{(1-x)(1-y)(1-z)}{(1-p)(1-q)}.
\end{equation}

\begin{table}[t]
\centering
\begin{tabular}{|c || c | c | c | c || c |} 
 \hline
  & $\Delta$ & $j_1$ & $j_2$ & $q_i$ & $f$ \\
 \hline\hline
 $X_l$ & $1$ & $0$ & $0$ & $\delta_{il}$ & $x, y, z$ \\
 \hline
 $\psi_l$ & $\frac{3}{2}$ & $\frac{1}{2}$ & $0$ & $\frac{1}{2} - \delta_{il}$ & $- y z, - z x, - x y$ \\ 
 \hline
 $\lambda_\pm$ & $\frac{3}{2}$ & $0$ & $\pm\frac{1}{2}$ & $\frac{1}{2}$ & $-p,-q$ \\ 
 \hline
 $F$ & $2$ & $1$ & $0$ & $0$ & $x y z = p q$ \\ 
 \hline
 $\partial \lambda = 0$ & $\frac{5}{2}$ & $\frac{1}{2}$ & $0$ & $\frac{1}{2}$ & $p q$ \\ 
 \hline
 $\partial_\pm$ & $1$ & $\frac{1}{2}$ & $\pm \frac{1}{2}$ & $0$ & $p, q$ \\
 \hline
\end{tabular}
\caption{Letters and their charges in the index of $\mathcal{N}=4$ Yang-Mills. We abbreviated the scalar fields $X$,$Y$,$Z$ as $X_l$ and chiralinos $\psi_X$,$\psi_Y$,$\psi_Z$ as $\psi_l$. The fields are charged under angular momenta $U(1)_{1} \times U(1)_{2} \subset SO(4)$ and R-symmetries $U(1)_i^3 \subset SO(6)$. In the free limit of $\mathcal{N}=4$ SYM, there is also a bonus $U(1)$ symmetry $Y$ \cite{Intriligator:1998ig,Intriligator:1999ff,Chang:2013fba} which assigns $1$ to all fields and the supercharge $Q$ and assigns $0$ to derivatives.}.
\label{table: N=4 YM fields}
\end{table}

It has been recently demonstrated that the $1/16$-BPS index of $U(N)$ $\mathcal{N}=4$ SYM reproduces the growth of the entropy of $1/16$-BPS black holes in $\AdS_5$, via direct numerical studies of the index at finite $N$ \cite{Murthy:2020rbd,Agarwal:2020zwm}. $1/16$-BPS black holes possess large angular momenta and large R-charges \cite{Gutowski:2004ez,Gutowski:2004yv,Chong:2005hr,Kunduri:2006ek,Boruch:2022tno}, where our angular momenta are related to those in the $\AdS_5$ black hole literature as $j_{1,2} = \frac{1}{2} (J_1 \pm J_2)$. In studies of the index, this feature of $1/16$-BPS black holes is often implemented at the level of fugacities by the scaling
\begin{equation} \label{eq: N=4 w scaling}
    x = y = z = w^2, \quad p = q = w^3
\end{equation}
consistent with $x y z = p q$.

It was also recently proposed \cite{Imamura:2021ytr} that the $1/16$-BPS index of $\mathcal{N}=4$ super Yang-Mills can be expressed as an expansion in terms of sums over stacks of three giant gravitons that wrap independent $S^3$ in $S^5$.\footnote{A similar proposal was also made by the author and D. Gaiotto in \cite{Gaiotto:2021xce} for the $1/16$-BPS index in the regime where one R-charge is taken to be large while other angular momenta and charges were held finite. In that regime, it is sufficient to consider the expansion with respect to a single giant. The expressions for the giant graviton indices were derived by considering fluctuation modes of a probe giant in supergravity in \cite{Imamura:2021ytr} and by considering modifications of determinants in gauge theory in \cite{Gaiotto:2021xce}.} They studied the index in the scaling \eqref{eq: N=4 w scaling} where $1/16$-BPS black holes should be relevant and showed that giant graviton contributions reproduce the finite $N$ gauge theory index up to a third set of corrections that enter at $O(w^{2( 3N + 3^2)})$ in the expansion
\begin{equation}
    Z_N = Z_\infty \sum_{k_X,k_Y,k_Z} x^{k_X N} y^{k_Y N} z^{k_Z N} \hat{Z}_{(k_X,k_Y,k_Z)}.
\end{equation}

We examine this proposal in light of the findings of this work. Maximal giant gravitons in IIB string theory on $\AdS_5 \times S^5$ wrap a maximal $S^3$ in $S^5$. They are dual to determinant operators $\det X$, $\det Y$, $\det Z$ of complex scalars in the field theory. The scalars possess R-charges but no angular momenta, so these giant gravitons start out with $N$ units of an R-charge and no Lorentz angular momentum. The adjoint single-letter index of $(\det X)^k$ is
\begin{equation} \label{eq: N=4 adjoint letter X}
    \hat{f}_X^X = \frac{1}{(1-y)(1-z)} \left( p + q - p q + \frac{1}{x} - \frac{p}{x} - \frac{q}{x} + \frac{p q}{x} - y - z + y z \right).
\end{equation}
Positive terms in \eqref{eq: N=4 adjoint letter X} become denominators in the gauge integrand. They suggest the following modifications
\begin{align}
    X &\to Y^m Z^l \ \partial_+ X \nonumber \\
    X &\to Y^m Z^l \ \partial_- X \nonumber \\
    X &\to Y^m Z^l \ \nonumber \\
    X &\to Y^m Z^l \ \lambda_+ \lambda_- \nonumber \\
    X &\to Y^m Z^l \ F
\end{align}
for the giant. In the $U(k)$ worldvolume theory dual to $(\det X)^k$, it is apparent that scalars $Y$ and $Z$ assume the role of BPS derivatives. They can supply large R-charges $q_2$,$q_3$ for the giant. Open strings can also source angular momentum via $\partial_\pm$ and $F$, but it would take many such excitations for the giant to acquire angular momenta that scale with its R-charge. One can check that bifundamental excitations between scalar determinants do not supply any more angular momenta.

The puzzle in the $1/16$-BPS sector is whether additional branes dual to determinants that are not of the form $\det X$, $\det Y$, $\det Z$ are necessary. Our prescription suggests that there could be more determinants at play. Operators such as $\det F$, $\det \psi_X^2$, $\det \psi_X \psi_Y$, or $\det \partial_+ \partial_- F$ do not have zeromodes and can directly supply the large angular momenta expected of $1/16$-BPS black holes. These determinants have large $j_1$ (i.e. large $J_{1,2}$) in $\AdS_5$. If additional branes are not necessary, it means that a collection of ordinary giant gravitons can acquire large AdS angular momenta comparable to that of a BPS black hole, purely through adjoint open string excitations, at energies of $O(N^2)$.

We computed the brane index up to the third set of corrections
\begin{equation}
    \sum_{k_X + k_Y + k_Z = 3} x^{k_X N} y^{k_Y N} z^{k_Z N} \hat{Z}_{(k_X, k_Y, k_Z)}
\end{equation}
in the scaling \eqref{eq: N=4 w scaling} at small $N$. Interestingly, we find that additional contributions due to $\det F$ or $\det \lambda_+ \lambda_-$, with bare fugacities $(x y z)^N$ and $(p q)^N$, are \textit{not} necessary to reproduce the gauge theory index up to the order checked.

One needs to compute higher corrections from giant graviton indices to see whether other determinants such as $\det \psi_X^2$, $\det \psi_X \psi_Y$, or $\det \partial_+ \partial_- F$ contribute. However, higher corrections are difficult to compute directly. It would be fruitful to develop an approach to classify the contributing determinant operators given the single-letter index of a $U(N)$ gauge theory.

\section{Structure of bulk microstates} \label{sec: structure of bulk microstates}

In various examples, we constructed the indices of worldvolume gauge theories of bulk branes using the data of a dual $U(N)$ gauge theory. The bulk microstates are configurations of strings and branes whose interactions are organized in terms of a quiver diagram.

While we think the microstates themselves are best interpreted in the tensionless limit of string theory, their collective degeneracies are meaningful in supergravity because indices are protected under changes in $\lambda$. This section is driven by two questions regarding the organization of microstates in the bulk: (1) How do the microstates assemble in the BPS Hilbert space to give results consistent with gauge theory and supergravity? (2) In what circumstances can we see degeneracies consistent with highly excited geometries like black holes? For the remainder of this section, we assume that $N$ is large but finite.

\subsection{A brane number grading} \label{subsec: brane number grading}

Our proposal for the microstates was based on gauge theory determinant operators that are necessary in the brane expansion for superconformal indices. The brane expansion tells us that the gauge/string theory index is a coherent sum over stacks of branes and their excitations.

Being the index of a $\prod_i U(k_i)$ worldvolume theory, each brane index at large $k_i$ should exhibit growth that is at least comparable to the growth of the $U(N)$ gauge theory index. That is, if the $U(N)$ gauge theory admits BPS black hole states, then each $\prod_i U(k_i)$ worldvolume theory at large $k_i$ may also have black hole states. A natural question is, how can an infinite sum over brane indices, each of which has black hole growth, result in the degeneracies of a single $U(N)$ gauge theory?

At first blush, one may attribute these cancellations to the fact that we are computing an index. However, we noticed in Section \ref{subsec: half-bps} that such cancellations occur even for BPS partition functions. We argue that cancellations of coefficients occur in the bulk string theory in a way that cannot purely be attributed to the operator $(-1)^F$ in the index.

We recapitulate the observation regarding the half-BPS partition function of $\mathcal{N}=4$ super Yang-Mills. There are no fermions in this sector of gauge theory, so the index is equal to the BPS partition function. The partition function of $k$ giant gravitons is
\begin{equation} \label{eq: half-BPS giant graviton part fn}
    \hat{Z}_k = \prod_{m=1}^k \frac{1}{1 - x^{-m}} =  \frac{(-1)^k x^{k(k+1)/2}}{\prod_{m=1}^k (1 - x^{m})}.
\end{equation}
Power expanding in $x^{-1}$ is natural for the spectrum of maximal giants on $S^3 \subset S^5$: open string excitations can only take away $U(1)_X$ R-charges because the giant cannot grow further. However, the $U(N)$ gauge theory sees physics as a power series in $x$: half-BPS excitations are multitraces $\prod_i \Tr X^{m_i}$ acting on the vacuum state. When comparing the brane and gauge theory partition functions, we should therefore analytically continue $x$ from outside the unit disk to inside the unit disk in $\mathbb{C}$.

An important consequence of the analytic continuation is that $\hat{Z}_k$ gets an overall minus sign for odd $k$. This effect would be most naturally explained if there is an additional $\mathbb{Z}_2$ grading in the half-BPS Hilbert space of IIB string theory. Let us suppose that the right-hand side of
\begin{equation}
    Z_N = Z_\infty \sum_{k=0}^\infty x^{k N} \hat{Z}_k
\end{equation}
is telling us how to compute the half-BPS partition function of IIB strings on $\AdS_5 \times S^5$ in terms of target space objects. Here we are not computing an index, but the analytic continuation still introduces negative signs for odd stacks of branes if we look at the spectrum as a series in $x$. The negative contributions could be explained naturally if the string partition function \textit{computed as a series in $x$} contains an operator such as $(-1)^B$, where $B$ is a brane number operator for a particular set of branes about which the microstates organize. That is, $B$ is an operator measuring the RR charge of $k$ bulk giant gravitons. The operator $(-1)^B$ is unnecessary in a nonperturbative computation of the half-BPS partition function using the exact brane spectrum \eqref{eq: half-BPS giant graviton part fn}. It is nevertheless an emergent operator that must be taken into account when we compute the spectrum as a series in $x$.

The coefficients of the half-BPS partition function $Z_N$ are therefore graded dimensions of the Fock space of giant gravitons and their analytically-continued excitations. In the literature \cite{Lin:2004nb,Berenstein:2004kk,Dutta:2007ws}, D3 giant gravitons in the half-BPS sector have been described as holes in fermi droplets even though the dual CFT operators $\det X$ are bosonic. Negative signs coming from the analytic continuation offer a precise explanation as to why D3 giant gravitons behave effectively as fermions. It would be interesting if the notion of a brane number grading persists beyond protected sectors.

\begin{figure}[t]
\centering
\includegraphics[scale=0.45]{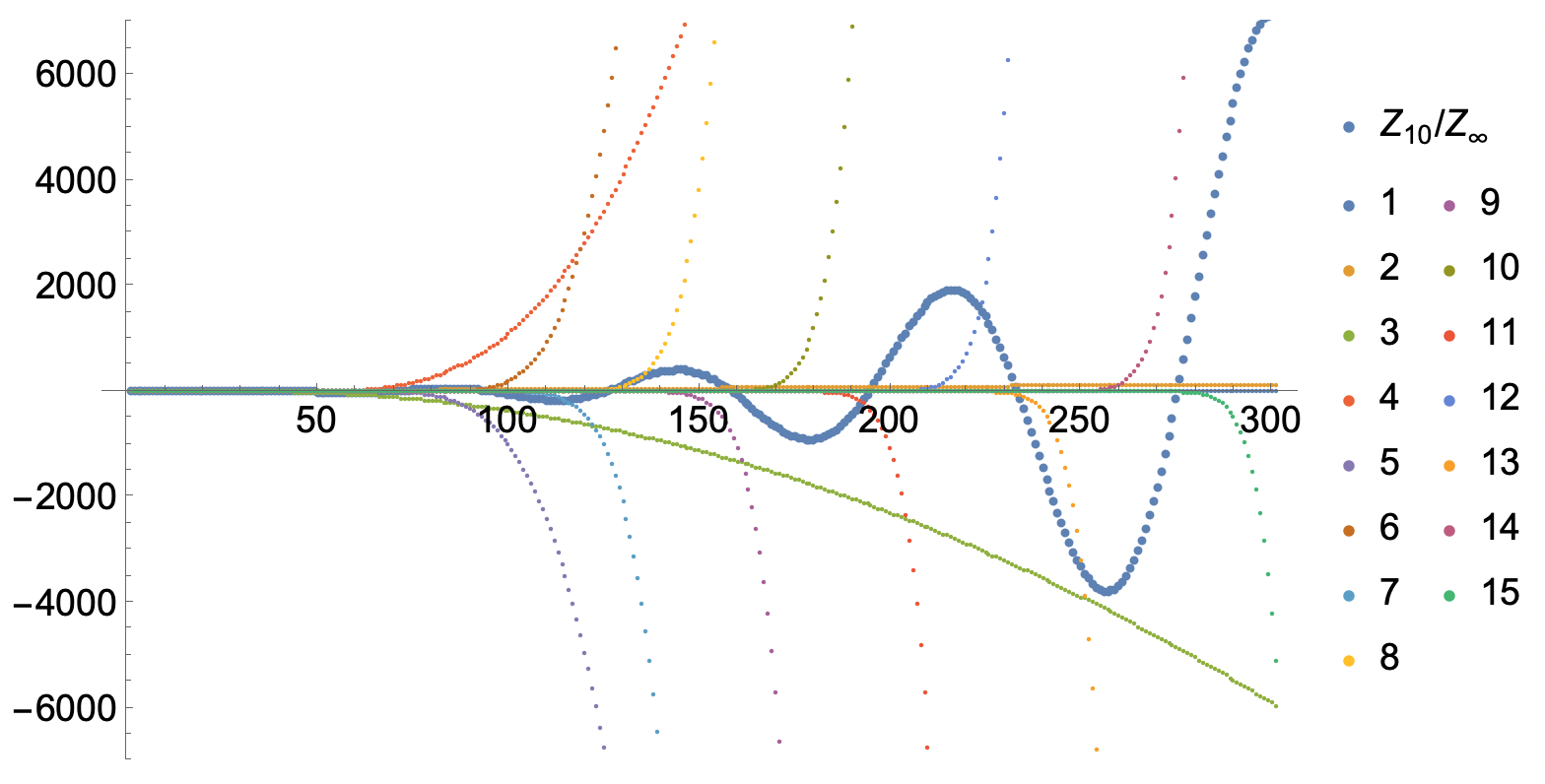}
\caption{Coefficients of the normalized half-BPS partition function $Z_N/Z_\infty$ with $N=10$ (large blue) versus charge $n$. Large cancellations between brane partition functions $x^{k N} \hat{Z}_k$ (in various colors for increasing $k$ up to $k=15$) induce oscillations that are roughly of order $N$. The normalized spectrum can have negative coefficients but the unnormalized half-BPS spectrum is always positive.}
\label{fig: half bps plot}
\end{figure}

In gauge theory, the effective grading is reflected in the truncation of single trace operators due to trace relations at finite $N$. The truncation affects the multitrace spectrum in charge intervals of order $\sim N$. In Figure \ref{fig: half bps plot}, we plot large cancellations between even and odd branes for the $1/2$-BPS partition function. The normalized spectrum $Z_N/Z_\infty$ shows oscillations of order $\sim N$ due to stacks of $k$ branes entering at every interval of $\sim N$. When one considers the log of absolute values of the coefficients in the normalized partition function, sinusoidal oscillations in the spectrum become bumps of order $N$ in the entropy as a function of charge $n$.\footnote{Oscillations in the spectral density of matrix models dual to low dimensional gravity were observed in \cite{Saad:2019lba,Johnson:2021zuo,Johnson:2022wsr}. Though they have a very different physical origin, it is interesting that the oscillations are signaling the presence of brane-like microstates there as well.} These bumps were observed directly for the $1/16$-BPS index in \cite{Murthy:2020rbd,Agarwal:2020zwm}. In particular, the authors of \cite{Agarwal:2020zwm} noticed that there are order $N$ fluctuations in the sign of index coefficients that are present on top of order $1$ sign fluctuations. On the string side, we think that order $N$ fluctuations are due to the presence of the effective operator $(-1)^B$, while order $1$ fluctuations are due to $(-1)^F$. An important problem would be to understand how the effective operator $(-1)^B$ arises directly in string theory.

\subsection{Black holes and wall-crossing} \label{subsec: black holes and wall-crossing}

While superconformal indices are protected against continuous deformations, their coefficients can exhibit discontinuous jumps across codimension-1 surfaces in the space of global symmetry fugacities. This phenomenon, in a more general sense, is known as wall-crossing \cite{Kontsevich:2008fj,Kontsevich:2009xt,Yamazaki:2010fz}. The surfaces divide the space of fugacities into various chambers; in each chamber, the index admits an invariant series expansion.

Let us take all global symmetry fugacities to lie inside the complex unit disk. Then the indices of $U(N)$ gauge theories in Section \ref{sec: examples} do not exhibit wall-crossing if we consider ``resolved'' fugacities. For example, for the twisted $\mathcal{N}=1$ and $\mathcal{N}=2$ gauge theories, we resolved the fugacity of the field strength component $F_{++}$ because it coincided with the fugacity for the gaugino equation of motion. In the $\mathcal{N}=2$ case, the resolution removed a pole of the $U(N)$ index at $t = p q$.\footnote{We emphasize that this is a pole of the unresolved $U(N)$ $\mathcal{N}=2$ gauge theory index and \textit{not} the pole that appears in individual brane indices at $x t = p q$.} \footnote{An alternative way to make the discussion in this section go through without resolving the fugacities by hand is to simply restrict the domain of fugacities, e.g. $\lvert \frac{p q}{t} \rvert < 1$ for the twisted $\mathcal{N}=2$ example.} The absence of poles in the space of global symmetry fugacities allows the gauge theory index to have a fixed charge lattice, where each point of the resolved charge lattice carries a definite number for the degeneracy of states.

On the other hand, brane indices do exhibit wall-crossing in their series coefficients because the individual brane indices have poles when one or more fugacities collide. When such poles are present, coefficients depend on the order in which fugacities are expanded.\footnote{A simple example is the expression $\frac{1}{x-y}$ which can be expanded in $x/y$ or $y/x$. The former is appropriate when $|x/y|<1$ and the latter when $|y/x|<1$. The expansions in $x$ and $y$ do not commute and there is a codimension-1 surface in the space of fugacities across which coefficients jump.} In examples such as the $1/4$-BPS sector of $\mathcal{N}=4$ SYM and the Higgs branch of M2 worldvolume theory (see Appendix \ref{app: m2 higgs branch}), it is possible to see the presence of these poles directly because there are exact formulas for brane indices.

The brane expansion relates two kinds of indices: a gauge theory index that does not have wall-crossing and brane indices that exhibit wall-crossing. We can ask what happens in string theory that renders the two pictures compatible.

For concreteness, consider the following $U(k)$ brane index that counts modifications of $(\det Y)^k$ in the $1/4$-BPS sector:
\begin{equation}
    \hat{Z}_{(0,k)} = \frac{\prod_{m = 1}^\infty (1 - x^m)}{\prod_{m=1}^k (1-y^{-m}) \prod_{m = 1}^\infty (1 - y^{-k} x^m)}.
\end{equation}
Suppose $|x|,|y|<1$. Then the index has poles at $y^k = x^m$ with $m=1,2,\cdots$. Let us fix $x$ to be some nonzero complex number inside the unit disk, and see what happens to the series in $y$. Poles in the $y$ plane are at $y = x^{m/k}$ and the sequence of poles approach zero as $m \to \infty$. Thus, the radius of convergence of a series in $y$ vanishes. The index $\hat{Z}_{(0,k)}$ goes to zero as $y \to 0$ with an appropriate continuation of the expression. Depending on what fugacity is expanded first, we see that entire brane contributions can vanish. If we now repeat the same exercise with fixed $y$, the series in $x$ has finite radius of convergence at $|x| = y^{k}$. The index will have nonzero series coefficients in $x$ and $y$ when expanded in that order but not the other way around.

The lesson from the $1/4$-BPS example extends to general situations. Brane indices have poles when fugacities approach rational powers of one another. The poles force the index coefficients to vanish unless the index is expanded in a particular fugacity order. Therefore, most brane contributions actually drop out when we expand the fugacities in some particular order. If fugacity $x$ is associated with a charge $\mathcal{C}_X$ that only acts on the scalar field $X$, the correct degeneracy of states can be reproduced just by the sum over brane indices of $(\det X)^k$ and no other branes. Situations where it was sufficient to consider only a single type of brane were explored in our previous work \cite{Gaiotto:2021xce}. In a microcanonical language, this situation corresponds to fixing all but one charges in a charge lattice and looking at the growth of degeneracies along the unfixed charge. Other charges can be fixed to be any value, so all degeneracies of the gauge theory index can be retrieved using an expansion involving one type of brane.

If we are interested in finding the set of bulk microstates that have a chance of forming a basis of the free BPS Hilbert space, rather than in reproducing degeneracies, then we need to find the set of branes whose indices will sum convergently to the gauge theory index via the brane expansion. In particular, the infinite set of poles between fugacities must cancel in the full sum. The cancellation of the poles allow all fugacities to take finite values. The combined brane indices become ``bulk'' invariants under wall-crossing. Invariance under wall-crossing in the space of fugacities was guaranteed by the resolved gauge theory index, but it is nice to see how the property manifests itself in the bulk. We think that the microstates proposed in Section \ref{sec: examples} form bulk invariants under wall-crossing phenomena that are present in individual brane indices.

We are finally ready to answer the question: in what situation will we find degeneracies consistent with highly excited geometries, such as BPS black holes or bubbling geometries that break more than half of supersymmetries? We claim that these degeneracies arise at the cancelled poles in the space of fugacities. Such was the limit where two or more fugacities become coincident or, more generally, become rational multiples of each other. In string theory, the interpretation is that different types of branes are forming bound states at the cancelled poles.

It is fascinating to see how degeneracies assemble when branes form bound states. At the poles, individual brane indices in the brane expansion diverge. When one sums over all terms that are relevant at a given charge, the divergences cancel among branes and yield integer coefficients. If we keep $N$ abstract when combining brane terms, we find interesting patterns in the degeneracy of states. Namely, the degeneracies become polynomials in $N$. In the coincident limit $x = y = w$ of the M2 Higgs branch index in Appendix \ref{app: m2 higgs branch}, we find
{\allowdisplaybreaks
\begin{align}
    \frac{Z_N}{Z_\infty} = &1 + w^N \Big[ (-N-2) w-2 (N+1) w^2+(2-4 N) w^3+(13-7 N) w^4+(40-12 N) w^5 + \cdots \Big] \nonumber \\
    &+ w^{2 N} \frac{1}{2!}\Big[ (-N^2-9 N-20) w^4-4 (N^2+8 N+17) w^5+(-13 N^2-89 N-180) w^6 \nonumber \\
    &\qquad \qquad -2 (17 N^2+95 N+186) w^7+(-81 N^2-347 N-694) w^8 + \cdots \Big] \nonumber \\
    &+ w^{3 N} \frac{1}{3!}\Big[ (N^3+24 N^2+191 N+504) w^{10}+6 (N^3+23 N^2+179 N+472) w^{11} \nonumber \\
    &\qquad \qquad +(26 N^3+570 N^2+4330 N+11418) w^{12}+6 (15 N^3+312 N^2+2313 N+6110) w^{13} \nonumber \\
    &\qquad \qquad +3 (91 N^3+1786 N^2+12921 N+34254) w^{14}+\cdots \Big] \nonumber \\
    &+ w^{4 N} \frac{1}{4!}\Big[ (N^4+50 N^3+935 N^2+7750 N+24024) w^{20} \nonumber \\
    &\qquad \qquad +8 (N^4+49 N^3+905 N^2+7469 N+23244) w^{21} \nonumber \\
    &\qquad \qquad +(43 N^4+2062 N^3+37589 N^2+308858 N+965208) w^{22} + \cdots \Big] + \cdots
\end{align}}

\noindent We do not expect to see black holes in the above sector, but there may be an interpretation for the growth of states in terms of bubbling geometries or complex saddles that break more than half of supersymmetries.

In the coincident limit
\begin{equation}
    x=y=z=w^2, \quad p=q=w^3
\end{equation}
of the $1/16$-BPS sector of $\mathcal{N}=4$ SYM, we find
{\allowdisplaybreaks
\begin{align}
    \frac{Z_N}{Z_\infty} = &1 + w^{2N} \frac{1}{2!} \Big[  (-N^2-5 N-6) w^2+2 (N^2+3 N+2) w^3-3 (N (N+1)) w^4+6 (N^2-N-2) w^5 \nonumber \\
    &\qquad \qquad +(-11 N^2+33 N+20) w^6+18 (N-5) N w^7-2 (14 N^2-98 N+57) w^8 + \cdots \Big] \nonumber \\
    &+ w^{4 N} \frac{1}{5!} \Big[ 2 (N^5+25 N^4+245 N^3+1175 N^2+2754 N+2520) w^8 \nonumber \\
    &\qquad \qquad -2 (8 N^5+165 N^4+1370 N^3+5775 N^2+12482 N+11160) w^9 \nonumber \\
    &\qquad \qquad +12 (6 N^5+105 N^4+760 N^3+2915 N^2+6054 N+5490) w^{10} \nonumber \\
    &\qquad \qquad -16 (16 N^5+235 N^4+1475 N^3+5180 N^2+10494 N+9765) w^{11} + \cdots \Big] \nonumber \\
    &+ w^{6 N}  \Big[ \frac{1}{8!}(- 3 N^8 - 180 N^7 - 4662 N^6 - 68040 N^5 - 611667 N^4 \nonumber \\
    &\qquad \qquad - 3466260 N^3 - 12084468 N^2 - 23681520 N -19958400) w^{18} + \cdots \Big] + \cdots
\end{align}}

\noindent The degree of the polynomial is related to the degree of the pole cancelled among the branes. We expect the pattern to persist more generally, where bound states of branes give a convergent expansion for the index that is, rather oddly, organized perturbatively in $N$ and nonperturbatively in $e^{-N}$. It would be interesting if a coherent sum over polynomials in $N$ of increasing degrees could produce degeneracies of highly excited geometries such as black holes.

\section{Discussion} \label{sec: discussion}

In this work, we studied how BPS microstates in string theory are organized in the dual gauge theory. We gave a prescription to construct the indices of string/brane microstate configurations from the gauge theory data. In various examples, we showed that the finite $N$ gauge theory index is a coherent sum of the indices of string/brane configurations. Finally, we discussed how the microstates assemble in the BPS Hilbert space and in what circumstances the branes can form bound states to produce black hole degeneracies.

We conclude with a brief list of open questions:
\begin{enumerate}
    \item Do the microstates, whose indices we compute, form a basis of the free BPS Hilbert space? It is possible that some branes are cancelled in the index. This may be possible to check by formulating a brane expansion for the free BPS partition function, whose existence was indicated in \cite{Murthy:2022ien}.
    \item Can we generalize our prescription to describe string duals of free $U(N)$ gauge theories in a non-BPS setting \cite{Gopakumar:2003ns,Gopakumar:2004qb,Gopakumar:2005fx,Aharony:2007fs,Chen:2022hbi}? It would be nice to make direct contact with tensionless worldsheet theories \cite{Gaberdiel:2021qbb,Gaberdiel:2021jrv}.
    \item How does the contributing set of determinants change when we change the basis of charges in the index? What kinds of determinants can appear? A mathematical study of the brane expansion would be helpful in classifying the contributing branes.
    \item It is curious that there is an effective grading in the bulk BPS Hilbert space at finite $N$ due to the analytic continuation of the brane partition function. It would be important to understand its origin directly in string theory.
    \item It would be interesting to apply our prescription to $U(N)$ gauge theories without known string duals. Generalizing our prescription to quiver gauge theories or to different gauge groups would allow one to study aspects of a range of string theories purely via the dual gauge theory. Some progress on single letter indices has been made in \cite{Arai:2019aou,Arai:2019wgv,Arai:2019xmp} from the bulk perspective, but a full definition of brane indices for these generalizations remains unknown.
    \item It would be important to better understand the connection between our string/brane configurations and highly excited supergravity solutions \cite{Callan:1988hs,Horowitz:1996nw,Horowitz:1997jc,Chen:2021dsw,Eberhardt:2020bgq,Eberhardt:2021jvj}.
\end{enumerate}

\subsection*{Acknowledgements}

We thank Nathan Benjamin, Kasia Budzik, Chi-Ming Chang, Kevin Costello, Pieter-Jan De Smet, Davide Gaiotto, Jaume Gomis, Ying-Hsuan Lin, Sameer Murthy, Eric Perlmutter, Surya Raghavendran, and Xi Yin for discussions. We are especially grateful to Davide Gaiotto for initial discussion, many valuable conversations, and feedback on the draft. We would like to thank Nathan Benjamin and Ying-Hsuan Lin in particular for discussions regarding the subject of Section \ref{sec: structure of bulk microstates}.

J.H.L. is supported by the Perimeter Institute for Theoretical Physics and in part by the NSERC Discovery Grant program. Research at Perimeter Institute is supported in part by the Government of Canada through the Department of Innovation, Science and Economic Development Canada and by the Province of Ontario through the Ministry of Colleges and Universities. This work was partly done at the Aspen Center for Physics, which is supported by National Science Foundation grant PHY-1607611.

\appendix

\section{Half-BPS brane index} \label{app: derivation half-bps}

We provide an explicit derivation of the brane index in the half-BPS sector, following \cite{Gaiotto:2021xce}. We assume that $N$ is large during the derivation, but the final result is exact at finite $N$.

In $\mathcal{N}=4$ SYM, determinant operators have dimension $N$ and are known to be dual to D3 giant graviton branes that wrap $\mathbb{R} \times S^3 \subset \AdS_5 \times S^5$. Finite modifications of the determinants correspond to open string excitations of these D3 branes.

We start with a single determinant
\begin{equation}
    \det X = \frac{1}{N!} \epsilon^{i_1 i_2 \cdots i_N} \epsilon_{j_1 j_2 \cdots j_N} X_{i_1}^{j_1} X_{i_2}^{j_2} \cdots X_{i_N}^{j_N}
\end{equation}
of fugacity $x^N$. The determinant can be modified by replacing a finite number of above $X$s by strings of letters in the theory. For example, we can replace an $X$ with $L_1 L_2 L_1$ to get
\begin{equation}
    \epsilon^{i_1 i_2 \cdots i_N} \epsilon_{j_1 j_2 \cdots j_N} (L_1 L_2 L_1)_{i_1}^{j_1} X_{i_2}^{j_2} \cdots X_{i_N}^{j_N}.
\end{equation}
There are some redundancies with counting such modifications, though. An example is a replacement of the form $X \to X W$ or $X \to W X$, where antisymmetry allows one to write $\text{Tr} W \det X$. We will interpret these redundancies as closed strings of $Z_\infty$ and include them back later. For now, let's focus on the nonredundant part that corresponds to open strings.

A helpful reformulation of the problem is to write the determinant as an integral over auxiliary (anti)fundamental fermions
\begin{equation}
    \int d\bar{\psi} d\psi \ e^{\bar\psi X \psi}.
\end{equation}
Introducing (anti)fundamental fermions adds ``boundaries'' to the 't Hooft ribbon diagrams, which helps understand why determinants should correspond to D-brane insertions. In terms of the fermion integral, determinant modifications correspond to operator insertions
\begin{equation}
    \int d\bar{\psi} d\psi \ e^{\bar\psi X \psi} (\bar\psi W_1 \psi) (\bar\psi W_2 \psi) \cdots
\end{equation}
where an open string excitation looks like $\bar\psi L_1 \cdots L_s \psi$. The redundancy mentioned above becomes a Ward identity for the fermions
\begin{equation}
    \int d\bar{\psi} d\psi \ e^{\bar\psi X \psi} ( \bar\psi X W \psi ) (\bar\psi W_1 \psi) \cdots = \int d\bar{\psi} d\psi \ e^{\bar\psi X \psi} \bigg( \frac{d}{d\psi} ( W \psi) \bigg) (\bar\psi W_1 \psi) \cdots.
\end{equation}

We can implement the Ward identities by introducing bosonic antifields $u, \bar{u}$, as well as a BRST differential $\delta$ acting as
\begin{align}
    \delta X  = \delta \psi &= \delta \bar\psi = 0 \nonumber \\
    \delta u &= X \psi \nonumber \\
    \delta \bar{u} &= \bar{\psi} X.
\end{align}
For proper counting, we will need to posit that $\delta$ is an extra part of the cohomological supercharge $Q$ that acts nontrivially only on the antifields $u, \bar{u}$. We assign ghost numbers $-1$ to $u,\bar u$, $+1$ to $\delta$, and $0$ to other fields. We are interested in operators in the BRST cohomology with ghost number $0$. The action of $\delta$ were written so that redundancies due to replacements $X \to X W $ or $X \to W X $ become $\delta$-exact.

The auxiliary fundamental and anti-fundamental letters are counted by the single letters $v=(x-1)\sigma$ and $\bar{v} = (1-x^{-1})\sigma^{-1}$. $\lambda$ denotes a fugacity for an extra $U(1)$ symmetry which only acts on these auxiliary variables. It will drop out of calculations now but will be useful soon.

One problem with this approach is that there is cohomology in non-zero ghost number. The operator $\bar{\psi} X \psi$ can come from either $\delta( \bar{\psi} u )$ and $\delta( \bar{u} \psi )$, so the combination $\bar{\psi} u + \bar{u} \psi$ will be $\delta$-closed but not exact. It gives a fermionic zeromode with ghost number $-1$ and trivial fugacity.   As this operator is the only problematic one, we will simply remove the fermion zeromode by hand in our counting.

Let's now look back at the large $N$ gaussian index formula with (anti)fundamentals
\begin{equation}
    Z_\infty(y_i) = PE\Big[ \frac{v \bar{v}}{1 - f} \Big] \prod_{n=1}^\infty \frac{1}{1 - f(y_A^n)}.
\end{equation}
The infinite prefactor is the redundant closed string spectrum $Z_\infty$, which we ignore for now. The large $N$ formula suggests that the ``effective'' single letter index governing the determinant fluctuations is
\begin{equation}
    \hat{f} = 1 + \frac{v \bar{v}}{1 - f} = 1 - \frac{(1-x)(1-x^{-1})}{1-f}
\end{equation}
where the extra factor of $1$ cancels the zeromode.  
 
Therefore, modifications of a single determinant, with the redundant sector stripped off, are counted by the index
\begin{equation}
    \hat{Z}_1  = PE[\hat{f}].
\end{equation}
Let's apply this relation to the half-BPS example with $f=x$. We get
\begin{equation}
    \hat{f} = x^{-1}.
\end{equation}
This makes sense. Here, the only nontrivial operator is $X$, so any modification of the determinant would correspond to replacing $X$ by the identity $I$. This would take away a single power of fugacity $x$, thus the inverse. $\hat{Z}_1$ should then be interpreted as a $U(1)$ gauge theory index on the worldvolume of a single D3 giant graviton. $x$ is mapped to its inverse, because determinant modifications remove $X$s. 

It is straightforward to consider the modifications of $k$ determinants using the fermion description with $k$ flavors of fermions.
\begin{equation}
    (\det X)^k = \int d\bar{\psi} d\psi \ e^{ \bar\psi^\alpha X \psi_\alpha}.
\end{equation}
Insertions
\begin{equation}
    \int d\bar{\psi} d\psi \ e^{\bar\psi^\alpha X \psi_\alpha} ( \bar{\psi}^\beta W_1  \psi_\gamma) ( \bar{\psi}^\delta W_2  \psi_\epsilon)
\end{equation}
with different fermion indices now correspond to open strings stretched between different pairs of $k$ coincident giant graviton branes. The only difference from the previous case is that there is an emergent $U(k)$ gauge symmetry on the giant graviton worldvolumes that must be imposed. For proper counting, we should also subtract $k^2$ fermion zeromodes by hand.

Modifications of $k$ determinants in the half-BPS sector are described by the index
\begin{equation}
    \hat{Z}_k(x) = \frac{1}{k!} \oint \prod_a \frac{d\sigma_a}{2 \pi i \sigma_a} \prod_{a \neq b} (1 - \sigma_a \sigma_b^{-1}) \ \exp \left( \sum_{n=1}^{\infty} \frac{1}{n} x^{-1} \sum_{a,b} \sigma_a^n \sigma_b^{-n} \right),
\end{equation}
with $\hat{f}$ defined in the same way as in the $U(1)$ case. $\sigma_a$ are fugacities for an extra $U(N)$ symmetry that only act on the auxiliary variables. They dropped out for $U(1)$, but they become gauge fugacities for $U(k)$. Due to the inverse power of $x$, an analytic continuation is necessary to evaluate the index as a power series in $x$. We implement the analytic continuation through the prescription in Section \ref{subsec: analytic continuation}.

Let's put back in the closed string sector and the bare fugacity $x^{kN}$ for $k$ determinants. The half-BPS index of $k$ giant gravitons and their open/closed string excitations is
\begin{equation}
    x^{k N} Z_\infty(x) \hat{Z}_k(x).
\end{equation}
$x^{k N}$ is the bare fugacity of determinant $(\det X)^k$, $Z_\infty$ is the closed string spectrum, and $\hat{Z}_k(x)$ are analytically-continued open string excitations on $k$ coincident giant gravitons.

\section{Multivariate residues} \label{app: multivariate residues}

In single-variable complex analysis on $\mathbb{C}$, the residue at a pole is specified only by the location of the pole, up to sign from the orientation of the integration contour. For functions of several complex variables, say on $\mathbb{C}^K$ where $K = \sum_i k_i$, residues at a pole can take different values depending on the integration cycle. It is therefore necessary to specify the integration cycle as well as the location of relevant poles for a proper integral definition of $\hat{Z}_{(k_1, k_2, \cdots)}$. Fortunately, the integrand of the brane index belongs to a well-studied class of integrands in algebraic geometry. We review the theory of multivariate residues, following closely \cite{griffiths2014principles,Larsen:2017aqb}.

We consider meromorphic $K$-forms
\begin{equation}
    \omega = \frac{h(\sigma) d\sigma_1 \wedge \cdots \wedge d\sigma_K}{g_1(\sigma) \dots g_K(\sigma)},
\end{equation}
where $g(\sigma) = \left\{ g_1(\sigma), \dots, g_K(\sigma) \right\}$ and $h(\sigma)$ are holomorphic functions in the neighborhood of a given pole of $\omega$. We define a pole of $\omega$ as a point $\sigma_p \in \mathbb{C}^K$ where $g(\sigma)$ has an isolated zero. Without loss of generality, let us assume that $g$ has an isolated zero at the origin $\sigma_a = 0 \in \mathbb{C}^K$. The zero can be degenerate. The residue of $\omega$ at $\sigma_a = 0$ is defined as
\begin{equation}
    \underset{\sigma= 0; g}{\mathrm{Res}} \ \omega = \frac{1}{(2 \pi i)^K} \oint_{\Gamma_\epsilon} \frac{h(\sigma) d\sigma_1 \wedge \cdots \wedge d\sigma_K}{g_1(\sigma) \dots g_K(\sigma)},
\end{equation}
where $\Gamma_\epsilon$ is a real $K$-torus
\begin{equation}
    \Gamma_\epsilon = \{ \sigma \in \mathbb{C}^K : \left| g_a(\sigma) \right| =\epsilon \}
\end{equation}
oriented such that
\begin{equation}
    d(\arg g_1) \wedge \cdots \wedge d(\arg g_K) \geq 0.
\end{equation}
An important distinction from the single-variable residue theorem is that the integration cycle $\Gamma_\epsilon$ of multivariate residues now depend on the choice of denominator functions $g_a$. Different choices of $g_a$ correspond to selecting different integration cycles, which in turn yields different residues. There are standard algorithms to evaluate the residues in practice when $g_a, h$ are polynomials in $\sigma_a$. Important properties of the multivariate residue can be found in \cite{griffiths2014principles,Larsen:2017aqb}.

The integrand of $\hat{Z}_{(k_1, k_2, \cdots)}$ contains more than $K$ terms in the denominator, and in fact involves several infinite products in general. Therefore, there immediately arises the question of how to partition the denominators into $K$ factors $\{ g_1(\sigma), \dots, g_K(\sigma) \}$. In Section \ref{subsec: analytic continuation}, we propose a prescription for the determinant partition based on the analysis of physical fields in gauge theory that can ``modify'' a determinant operator. Examples of the prescription can be found in Section \ref{sec: examples}. We truncate the infinite products in the denominator during a series evaluation. Imposing a cutoff is fine if we want to evaluate the brane index in power series, because the coefficients of the power series stabilize with the cutoff. The full brane index $\hat{Z}_{(k_1, k_2, \cdots)}$ corresponds to the limit where the cutoffs are taken to infinity.

Let us define the Jacobian determinant
\begin{equation}
    J(p) = \det_{a,b} \left( \frac{\partial g_a}{\partial \sigma_b} \right) \bigg\rvert_{\sigma = p}.
\end{equation}
If $J(p) \neq 0$, the pole is nondegenerate and we can compute the residue at $\sigma = p$ directly using the formula
\begin{equation} \label{eq: jacobian residue}
    \underset{\sigma= 0; \ g}{\mathrm{Res}} \ \omega = \frac{h(p)}{J(p)}.
\end{equation}
If $J(p) = 0$, the pole is of higher order and degenerate. We can still evalute the degenerate residue directly using the following theorem \cite{griffiths2014principles,Larsen:2017aqb}:
\begin{theorem}
(Transformation formula). Let $g = \{g_1, \cdots, g_K \}$ and $G = \{ G_1, \cdots, G_K \}$ be holomorphic maps $g_a(\sigma), G_a(\sigma): \mathbb{C}^K \to \mathbb{C}$ with $g^{-1}(0) = G^{-1}(0) = p$. Furthermore, suppose that
\begin{equation}
    G_a(\sigma) = \sum_{b=1}^K A_{a b}(\sigma) g_b(\sigma)
\end{equation}
for $A_{a b}(\sigma)$ holomorphic matrix components of $A$. That is, functions in the set $g$ and $G$ form zero-dimensional ideals (i.e. equations $g_a(\sigma) = G_b(\sigma) = 0$ are satisfied by a finite number of points $\sigma \in \mathbb{C}^K$) such that the ideals satisfy
\begin{equation}
    \langle g_1, \cdots, g_K \rangle \subseteq \langle G_1, \cdots, G_K \rangle.
\end{equation}
Then the residue at $p$ is
\begin{equation}
    \underset{\sigma= p; \ g}{\mathrm{Res}} \ \left( \frac{h(\sigma) d\sigma_1 \wedge \cdots \wedge d\sigma_K}{g_1(\sigma) \dots g_K(\sigma)} \right) = \underset{\sigma= p; \ G}{\mathrm{Res}} \ \left( \frac{h(\sigma) \det A(\sigma) \ d\sigma_1 \wedge \cdots \wedge d\sigma_K}{G_1(\sigma) \dots G_K(\sigma)} \right).
\end{equation}
\end{theorem}

The idea is to choose $G$ such that the functions $G_a$ are univariate:
\begin{equation} \label{eq: transformation formula}
    G = \left\{ G_1(\sigma_1), G_2(\sigma_2), \cdots, G_K(\sigma_K) \right\}.
\end{equation}
The multivariate residue then factorizes into a product of ordinary univariate residues. We refer the readers to \cite{Larsen:2017aqb} for explicit examples of computation using the transformation formula.

We claimed in Section \ref{subsec: analytic continuation} that all the relevant poles of the integrand condense at $\sigma = 0$. In fact, $\sigma = 0$ is the only common zero of the elements of $g$. In theory, the residue at this degenerate pole can be computed directly given the partition $g$ using the tranformation formula \eqref{eq: transformation formula}. In practice, the procedure is computationally intensive.

If we restrict to $U(N)$ gauge theories with only adjoint fields, we can simply the computation somewhat because a $U(1)$ gauge fugacity decouples from the rest of $U(N)$ gauge fugacities. Namely, let us set $\sigma_K = 1$ in the integrand and define
\begin{align}
    \tilde{\sigma} &= \{ \sigma_1, \cdots, \sigma_{K-1} \} \nonumber \\
    \tilde{g} &= \{ g_1(\tilde{\sigma}), \cdots, g_{K-1}(\tilde{\sigma}) \} \nonumber \\
    \tilde{h}(\tilde{\sigma}) &= \frac{h(\tilde{\sigma})}{g_K(\tilde{\sigma})}.
\end{align}
Then the degenerate pole at $\sigma = 0$ resolves into a (mostly) nondegenerate set of poles:
\begin{equation}
    \underset{\sigma= 0; \ g}{\mathrm{Res}} \ \left( \frac{h(\sigma) d\sigma_1 \wedge \cdots \wedge d\sigma_K}{g_1(\sigma) \dots g_K(\sigma)} \right) = \sum_{\tilde{p}} \underset{\tilde{\sigma} = \tilde{p}; \ \tilde{g}}{\mathrm{Res}} \ \left( \frac{\tilde{h}(\tilde{\sigma}) d\sigma_1 \wedge \cdots \wedge d\sigma_{K-1}}{g_1(\tilde{\sigma}) \dots g_{K-1}(\tilde{\sigma})} \right)
\end{equation}
The resolved poles $\tilde{p}$ are the set of points in $\mathbb{C}^{K-1}$ at which $(K-1)$-elements of $\tilde{g}$ have common zeroes. Most poles become nondegenerate and their residues can be computed directly via the Jacobian formula \eqref{eq: jacobian residue}. There can still be a degenerate pole at $\tilde{\sigma} = 0$, but its residue is much easier to compute than the original one at $\sigma=0$.

The above simplification does not occur in general. For $U(N)$ gauge theories with (anti)fundamental fields (see Appendix \ref{app: m2 higgs branch}), there is no decoupling of a gauge fugacity. However, our prescription for the integration cycle is still well-defined for evaluation in power series.

\section{Example: M2 Higgs branch theory} \label{app: m2 higgs branch}

We now consider the index counting Higgs branch operators of the worldvolume theory on $N$ coincident M2 branes \cite{Aharony:2008ug,Okazaki:2019ony,Crew:2020psc}. We focus on the UV description in terms of 3d ADHM $\mathcal{N}=4$ super Yang-Mills theory with $N_f = 1$, which consists of the following 3d $\mathcal{N}=4$ multiplets: an adjoint vectormultiplet, an adjoint hypermultiplet, and a (anti)fundamental hypermultiplet. The superconformal index for this theory involves a sum over monopole sectors, but the Higgs branch index receives contributions only from the zero monopole sector.

Here, our focus is to give independent definitions for brane indices $\hat{Z}_{(k_X,k_Y)}$ based on Section \ref{sec: organizing principles}. This demonstrates the applicability of our prescription to $U(N)$ gauge theories with adjoint and (anti)fundamental fields, for which no index generating functions are known in general. 

Higgs branch operators that contribute to the gauge theory index are hypermultiplet scalars $X,Y$ with a superpotential constraint. Scalars $X,Y$ transform with charges $(H,f)=(1,1),(1,-1)$ under ``Higgs'' and flavor symmetries $U(1)_H \times U(1)_f$. It will be useful to change the basis of charges to $(\mathsf{X},\mathsf{Y}) = \frac{1}{2}(H+f,H-f)$ for R-symmetries $U(1)_X \times U(1)_Y$. In the new basis, $X,Y$ have charges $(1,0), (0,1)$. The superpotential constraint $X \cdot Y = 0$ has $(1,1)$ and contributes to the counting with opposite sign.

\begin{figure}[t]
\centering
\includegraphics[scale=0.4]{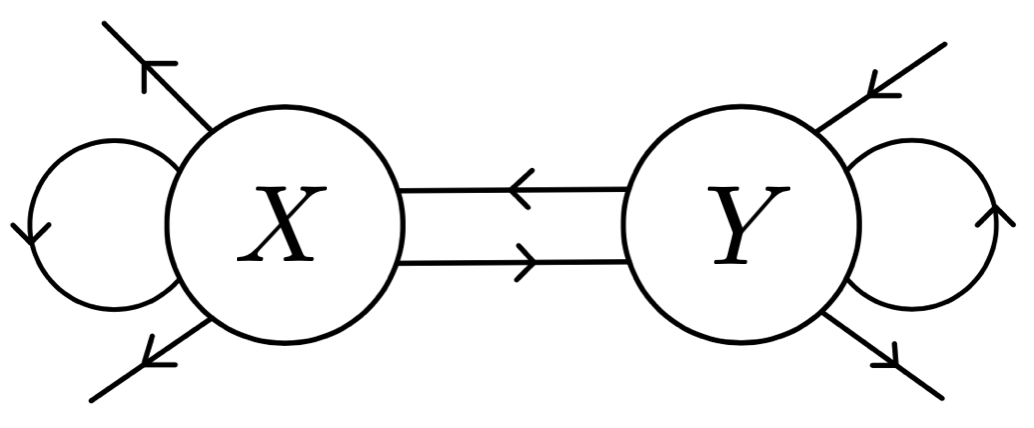}
\caption{Interactions of M5 giant graviton branes $\det X$ and $\det Y$.}
\label{fig: higgs quiver}
\end{figure}

The M2 Higgs branch adjoint and (anti)fundamental single-letter indices are
\begin{align}
    f &= x + y - x y \nonumber \\
    v &= x y \nonumber \\
    \bar{v} &= 1.
\end{align}
In the above, we shifted all gauge fugacities by $\sqrt{x y}$ for later convenience, resulting in $v \to v \sqrt{x y}$ and $\bar{v} \to \bar{v}/\sqrt{x y}$. The adjoint part is the same as that for the $1/4$-BPS index of $\mathcal{N}=4$ SYM, but the presence of (anti)fundamental parts will prevent the factorization of gauge nodes in the brane quiver of Figure \ref{fig: higgs quiver}.

The Higgs branch index and its relation via mirror symmetry to the Coulomb counterpart was discussed in \cite{Gaiotto:2021xce}. There, a generating function for Higgs/Coulomb branch indices
\begin{equation} \label{eq: Higgs generating fn}
    \mathcal{Z}(\zeta;x,y) = \sum_{N=0}^\infty \zeta^N Z_N(x,y) = \prod_{n,m=0}^\infty \frac{1}{1 - \zeta x^n y^m}
\end{equation}
was used to derive the giant graviton expansion
\begin{align} \label{eq: Higgs GGE}
    Z_N(x,y) &= Z_\infty \sum_{k_X,k_Y = 0}^\infty \frac{x^{k_X N} y^{k_Y N}}{\prod_{n=1}^{k_X} \prod_{m=1}^{\infty} (1 - x^{-n} y^m) \prod_{n=1}^{k_Y} \prod_{m=0}^{\infty} (1 - y^{-n} x^m) \prod_{n=1}^{k_X} \prod_{m=0}^{k_Y} (1 - x^{-n} y^{-m})} \nonumber \\
    &= Z_\infty \sum_{k_X,k_Y = 0}^\infty x^{k_X N} y^{k_Y N} \hat{Z}_{(k_X,k_Y)}
\end{align}
where
\begin{equation}
    Z_\infty = \prod_{n=1}^\infty \frac{1}{(1-x^n)(1-y^n)} PE\left[\frac{x y}{(1-x)(1-y)}\right] = \prod_{\substack{n,m=0 \\ (n,m) \neq (0,0)}}^\infty \frac{1}{(1-x^n y^m)}.
\end{equation}
Spurious poles that appear in each of $\hat{Z}_{(k_X,k_Y)}$ cancel when multiple such terms are combined. We expect $\hat{Z}_{(k_X,k_Y)}$ to be a reduced index for two intersecting stacks of $k_X$ and $k_Y$ M5 giant gravitons wrapping different $\mathbb{R} \times S^5 \subset \AdS_4 \times S^7$.

Single-letter indices for $\det X$ and $\det Y$ in the Higgs branch theory are
\begin{gather}
    \hat{f}_X^X = \frac{x^{-1} - y}{1 - y}, \qquad \hat{f}_Y^Y = \frac{y^{-1} - x}{1 - x}, \qquad \hat{f}^X_Y = y^{-1}, \qquad \hat{f}^Y_X = x^{-1} \nonumber \\
    \hat{v}^X = \frac{- x y}{1 - y}, \qquad \hat{\bar{v}}_X = \frac{- x^{-1}}{1 - y}, \qquad \hat{v}^Y = \frac{- x y}{1 - x}, \qquad \hat{\bar{v}}_Y = \frac{- y^{-1}}{1 - x}.
\end{gather}
From this, we readily write the brane index
\begin{align}
    \hat{Z}_{(k_X,k_Y)} = \frac{1}{k_X! k_Y!} \oint \prod_{a=1}^{k_X} &\frac{d\sigma_{a}^{X}}{2 \pi i \sigma_{a}^{X}} \prod_{c=1}^{k_Y} \frac{d\sigma_{c}^{Y}}{2 \pi i \sigma_{c}^{Y}} \prod_{a \neq b} (1 - \sigma^{X}_a/\sigma^{X}_b) \prod_{c \neq d} (1 - \sigma^{Y}_c/\sigma^{Y}_d) \nonumber \\
    &\prod_{a,b=1}^{k_X} \prod_{m=0}^\infty \frac{1 - y^{m+1} \sigma^{X}_a / \sigma^{X}_b}{1 - x^{-1} y^{m} \sigma^{X}_a / \sigma^{X}_b} \prod_{c,d=1}^{k_Y} \prod_{m=0}^\infty \frac{1 - x^{m+1} \sigma^{Y}_c / \sigma^{Y}_d}{1 - y^{-1} x^{m} \sigma^{Y}_c / \sigma^{Y}_d} \nonumber \\
    &\prod_{a=1}^{k_X} \prod_{c=1}^{k_Y} \frac{1}{1 - y^{-1} \sigma^{X}_a / \sigma^{Y}_c} \frac{1}{1 - x^{-1} \sigma^{Y}_c / \sigma^{X}_a} \nonumber \\
    &\prod_{a=1}^{k_X} \prod_{m=0}^\infty ( 1 - x y^{m+1} \sigma_a^X ) ( 1 - x^{-1} y^{m} / \sigma_a^X ) \nonumber \\
    &\prod_{c=1}^{k_{PQ}} \prod_{m=0}^\infty ( 1 - y x^{m+1} \sigma_c^Y ) ( 1 - y^{-1} x^{m} / \sigma_c^Y )
\end{align}
An important difference between this index and the $1/4$-BPS example is that the (anti)-fundamental terms supply an infinite number of additional poles at $\sigma_a^X = \sigma_c^Y = 0$. Because of this, the $U(1)$ factor of $U(N)$ no longer decouples and there is a nontrivial gauge integral even at rank one. These poles also result in a non-factorizing quiver for the system of string and brane configuations dual to the Higgs branch of M2 worldvolume theory.

In cases with (anti)fundamental fields, it is harder to determine whether the quiver diagram factorizes just by looking at the denominators. One can nevertheless write the denominator partition in the same manner as before. The (anti)fundamentals only contribute infinite factors of $\sigma_a^X, \sigma_c^Y$ so the partition $g$ is almost the same as the $1/4$-BPS case:
\begin{align}
    g_a^X &= (\sigma_a^X)^\infty \prod_{\substack{b=1 \\ b \neq a}}^{k_X} \prod_{m=0}^\infty (\sigma_a^X - x^{-1} y^m \sigma_b^X) \prod_{c=1}^{k_Y} (\sigma^{Y}_c - y^{-1} \sigma^{X}_a) (\sigma^{X}_a - x^{-1} \sigma^{Y}_c) \nonumber \\
    g_c^Y &= (\sigma_c^Y)^\infty \prod_{\substack{d=1 \\ d \neq c}}^{k_Y} \prod_{m=0}^\infty (\sigma_c^Y - y^{-1} x^m \sigma_d^Y),
\end{align}
where $a,b = 1,2, \cdots, k_X$ and $c,d = 1,2, \cdots, k_Y$. The infinite powers of gauge fugacities $\sigma_a^X, \sigma_c^Y$ are formal. Along with infinite products for other denominator factors, the powers are truncated appropriately in a power series computation.\footnote{For example, if $m$ is truncated at $m_0$, powers $(\sigma_a^X)^\infty, (\sigma_c^Y)^\infty$ should be truncated at $m_0+2$.} We checked the brane expansion up to total rank two, where there is indeed a nonvanishing mixed contributon $\hat{Z}_{(1,1)}$ as well as $\hat{Z}_{(2,0)}, \hat{Z}_{(0,2)}$.

\section{Checks for $\mathcal{N}=1$ vectormultiplet} \label{app: n1 yang mills checks}

We check the brane expansion for the twisted pure $\mathcal{N}=1$ Yang-Mills:
\begin{equation}
    Z_N = Z_\infty \sum_{k_F,k_\lambda = 0}^\infty u^{k_F N} (p q)^{k_\lambda N} \hat{Z}_{(k_F,k_\lambda)},
\end{equation}
where $\hat{Z}_{(k_F,k_\lambda)}$ contains all open string excitations and the sum is over all ranks of the gauge group $U(k_F) \times U(k_{\lambda})$. The background closed string spectrum is
\begin{equation}
    Z_\infty = \prod_{n=1}^\infty \frac{(1 - p^n)(1 - q^n)}{(1 - u^n)}.
\end{equation}
Let us scale the fugacities as
\begin{equation}
    p \to p w, \quad q \to q w, \quad u \to u w^2.
\end{equation}
and check the expansion in power series in $w$. The scaling satisfies the constraint $u = p q$. Other limits such as $p \to p w^2, q \to q w, u \to u w^3$ were checked as well.

In the $w$-scaling, the brane expansion organizes as
\begin{equation}
    \hat{G}_1 = u^{N} \hat{Z}_{(1,0)} + (p q)^{N} \hat{Z}_{(0,1)}
\end{equation}
at level 1,
\begin{equation}
    \hat{G}_2 = u^{2 N} \hat{Z}_{( 2, 0)} + (p q)^{2 N} \hat{Z}_{( 0,2)} + u^{N} (p q)^N \hat{Z}_{(1,1)} 
\end{equation}
at level 2, and so on. We show results at $N=2$. The gauge theory index normalized by the supergraviton spectrum is

\footnotesize
\begin{align}
    \frac{Z_2}{Z_\infty} - &1 = -p q w^4 (p q+u)-w^5 (p+q) (p^2 q^2-u^2)+w^6 (p^2 (2 q^2 u-q^4+u^2)+p^4 (-q^2)-p^3 q^3+u^2 (q^2-u)) \nonumber \\
    &+w^7 (-p-q) (p^2 (-q^2 u+q^4-u^2)+p^4 q^2-p^3 q^3+2 p q u^2-q^2 u^2)+w^8 (p^4 (q^4+u^2)-2 p^3 q^3 u+p^6(-q^2) \nonumber \\
    &-p^2 q^6+p q u^3+u^2 (q^4-u^2))+w^9 (-p-q) (p^3 (3 q^3 u-q^5+q u^2)+p^2 (q^6-4 q^2 u^2)+p^6 q^2 -p^5 q^3 \nonumber \\
    &-p^4 u^2+p q u^2 (q^2+u)-q^4 u^2)+w^{10} (p^6 (q^4+u^2)+p^4 q^2 (-q^2 u+q^4+u^2)-p^3 q u (-4 q^2 u+q^4+2 u^2) \nonumber \\
    &+p^2 (q^4 u^2-3 q^2 u^3-q^8+u^4)-p^5 q^3 (q^2+u)+p^8 (-q^2)+2 p q u^3 (u-q^2)+u^2 (q^2 u^2+q^6-u^3)) \nonumber \\
    &+w^{11} (-p-q) (p^5 q (q^4+u^2)+p^3 (4 q^3 u^2-q^7+q u^3)+p^2 (-q^2 u^3+q^8-u^4)-4 p^4 q^4 u+p^8 q^2-p^7 q^3 \nonumber \\
    &-p^6 u^2+p q u^2 (q^2 u+q^4-u^2)-q^2 u^4-q^6 u^2+u^5)+w^{12} (p^8 (q^4+u^2)-p^5 q (9 q^2 u^2-10 q^4 u+q^6+u^3) \nonumber \\
   & +p^4 (-16 q^4 u^2+6 q^2 u^3+5 q^6 u+q^8+2 u^4)+p^3 q^3 u (-9 q^2 u+q^4+9 u^2)-p^2 (3 q^6 u^2-6 q^4 u^3+q^2 u^4 \nonumber \\
   &+q^{10}+u^5)+p^7 (q^3 u-q^5)+p^6 q^2 u (5 q^2-3 u)+p^{10} (-q^2)+p (q u^5-q^5 u^3)+u^2 (2 q^4 u^2-q^2 u^3+q^8-u^4)) \nonumber \\
   &+O(w^{13}).
\end{align}
\normalsize

\noindent For the twisted $\mathcal{N}=1$ vectormultiplet in the $w$-scaling, power series for $\hat{G}_K$ start at $O(w^{2 K N})$. Subtracting corrections at level 1 leaves

\footnotesize
\begin{align}
    \frac{Z_2}{Z_\infty} - &(1 + \hat{G}_1) = p q w^8 (p^3 q^3-p q x^2-x^3)+w^9 (p+q) (p q-x) (p^3 q^3-p q x^2-x^3) \nonumber \\
    &+w^{10} (p^2+q^2-2 x) (p q-x) (p^3 q^3-p q x^2-x^3)+ w^{11} \frac{1}{p q} (p+q) (p q-x) (-p q x^3 (p^2-p q+q^2) \nonumber \\
    &+2 p^3 q^3 x^2+p^3 q^3 x (p^2+q^2)+p^4 q^4 (p-q)^2-p q x^4-x^5) +w^{12} (2 x^4 (3 p^2 q^2+p^3 q+p^4+p q^3+q^4) \nonumber \\
    &-p q x^3 (p^2+q^2) (p^2+p q+q^2)-p^2 q^2 x^2 (10 p^2 q^2+5 p^3 q+3 p^4+5 p q^3+3 q^4)+p^3 q^3 x (10 p^2 q^2 \nonumber \\
    &+5 p^3 q+p^4+5 p q^3+q^4)+p^4 q^4 (p-q)^2 (p^2+p q+q^2)-\frac{2 x^7}{p q}+2 p q x^5-2 x^6)+O(w^{13}).
\end{align}
\normalsize

\noindent Subtracting corrections at level 2 leaves

\footnotesize
\begin{align}
    \frac{Z_2}{Z_\infty} - &(1 + \hat{G}_1 + \hat{G}_2) = \nonumber \\
    &- w^{12} \frac{1}{x (x-p q)^2} (-3 p^6 q^6 x^2 (p+q)^2+p^5 q^5 x^4+p^5 q^5 x^3 (p^2+3 p q+q^2)-3 p^3 q^3 x^6 \nonumber \\
    &+p^7 q^7 x (3 p^2+5 p q+3 q^2)-p^8 q^8(p+q)^2+p q x^8)+O(w^{13}).
\end{align}
\normalsize

\noindent We expect corrections at level 3 to cancel the remaining term at $O(w^{12})$, and so on. Interestingly, we find that we do not need to impose $u = p q$ as long as the overall scaling by $w$ is consistent with the constraint.

\section{Checks for $\mathcal{N}=2$ vectormultiplet} \label{app: n2 yang mills checks}

The single-letter index for the twisted pure $\mathcal{N}=2$ Yang-Mills is
\begin{equation}
    f = \frac{- p - q + p q + x t + x - t}{(1-p)(1-q)}
\end{equation}
with the constraint $x t = p q$. The relevant letters and their charges are shown in Table \ref{table: pure N=2 YM fields}. Adjoint single-letter indices for dual branes that do not suffer from zeromodes are
\begin{align}
    \hat{f}_1 &= \frac{1}{1+t} \left[ p + q - p q + t + \frac{1}{x} - \frac{p}{x} - \frac{q}{x} + \frac{pq}{x} \right] \nonumber \\
    \hat{f}_{2} &= \frac{1}{1-x} \left[ p + q - p q + \frac{1}{t^2} - \frac{p}{t^2} - \frac{q}{t^2} + \frac{p q}{t^2} - \frac{1}{t} + \frac{p}{t} + \frac{q}{t} - \frac{pq}{t} + t - p t - q t + p q t - x \right] \nonumber \\
    \hat{f}_{3} &= \frac{1}{(1-x)(1+t)} \left[ 2 p + 2 q - \frac{1}{p} - \frac{1}{q} + \frac{1}{p q} - p q - p^2 q - p q^2 + p^2 q^2 + t - x - t x \right] \nonumber \\
    \hat{f}_{4} &= \frac{1}{(1-x)(1+t)} \left[ -1 + 2 p + 2 q - 2 p q - \frac{1}{t} + \frac{1}{p t} + \frac{q}{t} - \frac{q}{p t} + t + p t - p^2 t - p q t + p^2 q t - x - t x \right] \nonumber \\
    \hat{f}_{5} &= \frac{1}{(1-x)(1+t)} \left[ -1 + 2 p + 2 q - 2 p q - \frac{1}{t} + \frac{1}{q t} + \frac{p}{t} - \frac{p}{q t} + t + q t - q^2 t - p q t + p q^2 t - x - t x \right].
\end{align}
In shown order, the letters capture adjoint excitations of $\det X$, $\det \psi^2$, $\det \lambda_+ \lambda_-$, $\det \lambda_+ \psi$, and $\det \lambda_- \psi$. To avoid clutter, we numbered the determinants. The apparent fermion zeromodes in $\hat{f}_{4}$ and $\hat{f}_{5}$ cancel after one converts the single-letters into infinite products. There are also 20 bifundamental single-letter indices between the five gauge nodes that is simple to work out.

We explicitly check the brane expansion
\begin{equation}
    Z_N = Z_\infty \sum_{k_1,k_{2},k_{3},k_{4},k_{5}} x^{k_1 N} t^{2 k_{2} N} (p q)^{k_{3} N} (p t)^{k_{4} N} (q t)^{k_{5} N} \hat{Z}_{( k_1, k_{2}, k_{3}, k_{4}, k_{5} )}
\end{equation}
with
\begin{equation}
    Z_\infty = \prod_{n=0}^\infty \frac{(1-p^n)(1-q^n)}{(1-x^n)(1+t^n)}.
\end{equation}
Let us scale the fugacities as
\begin{equation}
    p \to p w, \quad q \to q w, \quad x \to x w, \quad t \to t w. 
\end{equation}
and check the expansion in power series in $w$. The scaling satisfies the constraint $x t = p q$.

In the $w$-scaling, the brane expansion organizes as
\begin{equation}
    \hat{G}_1 = x^{N} \hat{Z}_{( 1, 0,0,0,0 )}
\end{equation}
at level 1,
\begin{equation}
    \hat{G}_2 = x^{2 N} \hat{Z}_{( 2, 0,0,0,0 )} + t^{2 N} \hat{Z}_{( 0, 1,0,0,0 )} + (p q)^{N} \hat{Z}_{( 0, 0,1,0,0 )} + (p t)^{N} \hat{Z}_{( 0, 0,0,1,0 )} + + (q t)^{N} \hat{Z}_{( 0, 0,0,0,1 )}
\end{equation}
at level 2, and so on. We show results at $N=2$. The gauge theory index normalized by the supergraviton spectrum is

\footnotesize
\begin{align}
    \frac{Z_2}{Z_\infty} - &1 = w^3 (p-x) (x-q) (x-t)+w^4 (p^2 (-(q^2+t^2-x^2))-p t^2 (q+t-x)+(q-x) (x-t) (x (q+t) \nonumber \\
    &+t (q+t)+x^2)) +w^5 (-p^3 (q^2+t^2-x^2)-p^2 (q+t-x) (q^2-x (q+t)+q t+t^2)+p (-x^2 (q^2+4 q t+t^2) \nonumber \\
    &-2 q^2 t^2+x^3 (q+t)+3 q t x (q+t)+t^4)-t^2 x (t^2-2 q^2)-t^2 (q-t) (q+t)^2+q t x^3+x^2 (q-t)^2 (q+t)-x^5) \nonumber \\
    &+w^6 (-p^4 (q^2+t^2-x^2)-p^2 (q^3 t+q^4-2 q t^3+x^3 (q+t)-x (q-t) (q+t) (q+2 t)-t^2 x^2-2 t^4-x^4) \nonumber \\
    &+p^3 (x (q^2+q t+t^2)-(q+t) (q^2+t^2)) -p (q-x) (t-x) (-x^2 (q+t)-t (q+t) (3 t-q)+t^2 x+x^3) \nonumber \\
    &+x^2 (q^2 t^2+q^4+4 q t^3+t^4)-q^2 t^2 (q-t) (q+2 t)+q t^2 x (q-3 t) (q+t)-x^5 (q+t)+x^4(q+t)^2 \nonumber \\
    &-t x^3 (q+t)^2-x^6)+O(w^7).
\end{align}
\normalsize

\noindent For the $\mathcal{N}=2$ vectormultiplet in the $w$-scaling, power series for $\hat{G}_K$ start at $O(w^{K N})$ up to a constant shift. Subtracting corrections at level 1 leaves

\footnotesize
\begin{align}
    \frac{Z_2}{Z_\infty} - &(1 + \hat{G}_1) = w^4 \frac{1}{x} p q t  (x (p+q+t)-(p+t) (q+t)-x^2)+ w^5\frac{1}{x^2}( (x^3 (2 p^2 (q^2+q t+t^2)+p t (q+t) (2 \nonumber \\
    &q+t)+q t^2 (2 q+t))-x^2 (t^3 (2 p^2+3 p q+2 q^2)+2 p^2 q^2 t+p^2 q^2 (p+q)+t^4 (p+q)+t^2 (p+q)^3)+x \nonumber \\
    &(t^4 (p^2+3 p q+q^2)+p^2 q^2 t (p+q)+p^3 q^3+t^5 (p+q)+t^3 (p+q)^3+p q t^2 (p+q)^2)+p q t (p^2 \nonumber \\
    &(-(q^2+q t+t^2))-p t (q+t)^2-t^2 (q^2+q t+t^2))+x^5 (p (q+t)+q t)-x^4 (p+q+t) (p (q+t)+q \nonumber \\
    &t)))+w^6 (t^2 x^2 (p^2+3 p (q+t)+q^2+3 q t+t^2)-\frac{1}{x^3}(p q t (p+t) (q+t) (t^2 (p^2+p \nonumber \\
    &q+q^2)+p^2 q^2+t^4))+\frac{1}{x^2}(t^6 (p^2+3 p q+q^2)+t^5 (p+q) (p^2+3 p q+q^2)+t^4 (6 p^2 \nonumber \\
    &q^2+3 p^3 q+p^4+3 p q^3+q^4)+p^2 q^2 t^2 (p+q)^2+p^3 q^3 t (p+q)+p^4 q^4+t^7 (p+q)+p q t^3 (p+q)^3)+x (-t^3 \nonumber \\
    &(p^2+4 p q+q^2)+t^2 (p+q) (p^2+q^2)+p q t (p^2+3 p q+q^2)+p^2 q^2 (p+q)-3 t^4 \nonumber \\
    &(p+q)-t^5)-t^2 (p^2+p q+q^2)^2-2 p^2 q^2 t (p+q)-p^2 q^2 (p^2+p q+q^2)-\frac{1}{x}(t^3 (p+t) (q+t) (t \nonumber \\
    &(p+q)+(p+q)^2+t^2))+2 t^5 (p+q)+t^4 (2 p+q) (p+2 q)-t^3 (p-q)^2 (p+q)-x^5 (p+q+t) \nonumber \\
    &+x^4 (p+q+t)^2-x^3 (p+t) (q+t)(p+q+t)+t^6)+O(w^7).
\end{align}
\normalsize

\noindent Subtracting corrections at level 2 leaves

\footnotesize
\begin{align}
    \frac{Z_2}{Z_\infty} - &(1 + \hat{G}_1 + \hat{G}_2) = \nonumber \\
    &- w^6 \frac{p q t (-t^4 x+t^5+t x^4-x^5) (p^2 t (q t-x^2)+p (q^2 t^2+q x (-t^2-2 t x+x^2)+t x^3)+q t x^2 (x-q))}{x (p-t) (t-q) (p t-x^2) (q t-x^2)}+O(w^7).
\end{align}
\normalsize

\noindent We expect corrections at level 3 to cancel the remaining term at $O(w^6)$, and so on. Interestingly, we find that we do not need to impose $x t = p q$ as long as the overall scaling by $w$ is consistent with the constraint.

\bibliographystyle{JHEP}
\bibliography{ExactStringyMicrostates.bib}

\end{document}